\documentclass[%
reprint,
 amsmath,amssymb,
prb,
showkeys,
]{revtex4-2}

\usepackage{graphicx}
\usepackage{dcolumn}
\usepackage{bm}

\usepackage{enumitem} 
\usepackage{amsmath}
\usepackage{color}
\usepackage{soul}

\def\e{\begin{equation}}
\def\f{\end{equation}}
\def\_#1{{\bf #1}}

\def\.{\cdot}

\def\@#1{_{\rm #1}}

\begin{document}


\title{Emulating Bianisotropic Coupling Through Coherent Illumination}

\author{F. S. Cuesta}%
 \email{francisco.cuestasoto@aalto.fi}
\author{M. S. Mirmoosa}
\author{S. A. Tretyakov}%
\affiliation{%
 Department of Electronics and Nanoengineering, Aalto University\\
 P.O. Box 15500, Aalto FI-00076, Finland
}%

\date{\today}

\begin{abstract}
One of the main advantages of reciprocal bianisotropic metasurfaces is their capability to produce asymmetric scattering depending from which side they are illuminated and on the handedness of circularly polarized illuminations. For most applications, these metasurfaces are designed for illumination by a single source at a time. The resulting bianisotropic metasurface has a specific and usually complex geometrical structure that ensures the expected scattering produced under various illuminations. Here we show that geometrical asymmetry of metasurfaces can be emulated by using non-bianisotropic layers in presence of coherent illumination, which allows us to replicate and optically control the desired asymmetric scattering and chirality effects. 
In particular, the concept is developed on an example of emulating asymmetric scattering needed to create a 180$^\circ$ hybrid junction for plane waves. We show that this device can be realized either using a bianisotropic metasurface or a set of simple sheets with electric response under simultaneous illumination by two coherent waves.
\end{abstract}

\maketitle


\section{Introduction}

\begin{figure}[t]
  \centering
  \includegraphics[width=1\linewidth]{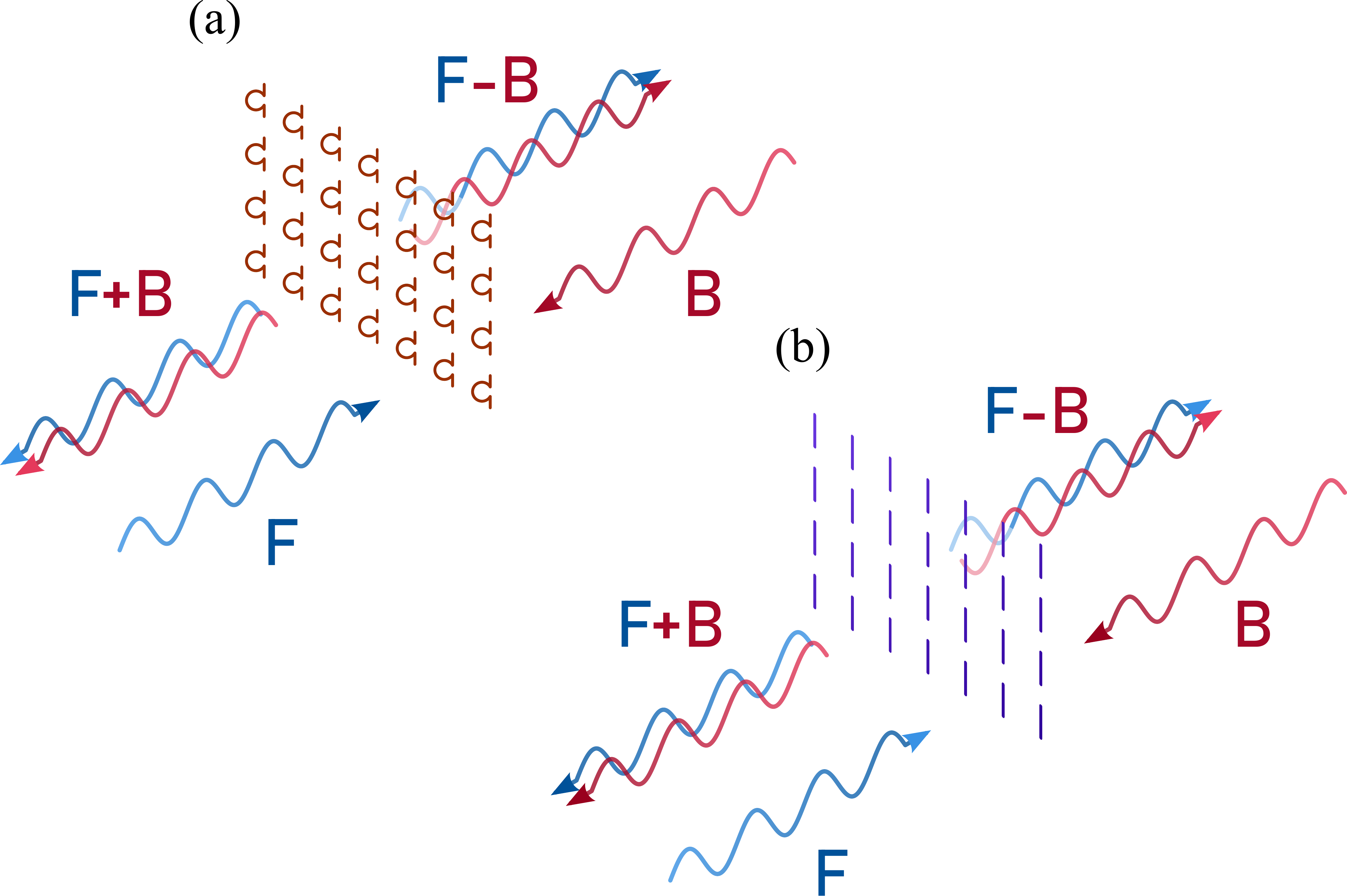}
  \caption{(a) A bianisotropic metasurface allows creation of asymmetric scattering that depends on the side where the source is located. Under illumination by two sources, such layer allows us to combine waves created by the two sources in different ways. (b) By exploiting the presence of two coherent sources, it is possible to replicate scattering of a bianisotropic layer. }
  \label{fig:delta_sigma_concept}
\end{figure}

Main interest in bianisotropic media and metasurfaces sprouts from their capability to produce cross-polarized (chiral) or asymmetric (omega) scattering response \cite{Serdyukov_2001_bianisotropic,Radi_2014_tayloring,Asadchy_2015_functional,Achouri_2016spatial,Asadchy_2018_bianisotropic}. In terms of metasurface design, bianisotropic scattering can be established using a single illumination source. 
If the metasurface has a linear response, its performance will be as expected whether it is illuminated from only one side at a time or simultaneously from both sides. As an example, with the use of bianisotropic inclusions it is possible to create a metasurface that allows combining two arbitrary incident waves so that the resulting scattered waves are equal to the sum and difference of the incident waves, as portrayed in Fig.~\ref{fig:delta_sigma_concept}(a). It is easy to see that such functionality requires bianisotropic response \cite{Radi_2014_tayloring,Achouri_2016spatial} (asymmetric reflection due to geometrical asymmetry of the metasurface).

Here, we ask ourselves the question if it is possible to emulate bianisotropic properties of metasurfaces using illumination of a simple non-bianisotropic metasurface by two or more coherent waves. The motivation is an analogy with other coherently-illuminated devices, like coherent absorbers \cite{coherent1,coherent2,Baranov_2017,coherent3,coherent4}. While perfect absorption of plane waves in thin layers requires excitation of both electric and magnetic surface currents in a finite-thickness layer \cite{absorbers}, using coherent illumination by two waves it is possible to realize full absorption in a sheet of negligible thickness that supports only electric surface current \cite{Zhang_2012_controlling}. The same conclusion is true for some other functionalities, such as retroreflection of plane waves \cite{retro}.
Based on these observations, we expect that also some asymmetric or polarization-sensitive reflection response can be achieved using non-bianisotropic metasurfaces, exploiting coherent illumination by two sources. By doing that, it becomes possible to emulate and optically control effects of complex bianisotropic layers using simple symmetric metasurfaces, as conceptualized in Fig.~\ref{fig:delta_sigma_concept}(b). We note that in paper \cite{coherent_chirality} it was shown that coherent illuminations can control the degree of optical activity in chiral layers. Here, we show that chiral and other bianisotropic effects can be emulated using completely nonchiral and symmetric structures. Actual realizations of coherent illuminations of metasurfaces usually use one source and various wave splitters or power dividers, see examples e.g. in Refs. \cite{Baranov_2017,coherent2,Zhang_2012_controlling,Shi,Rao,Thar}.

\begin{figure*}[t]
  \centering
  \includegraphics[width=0.85\linewidth]{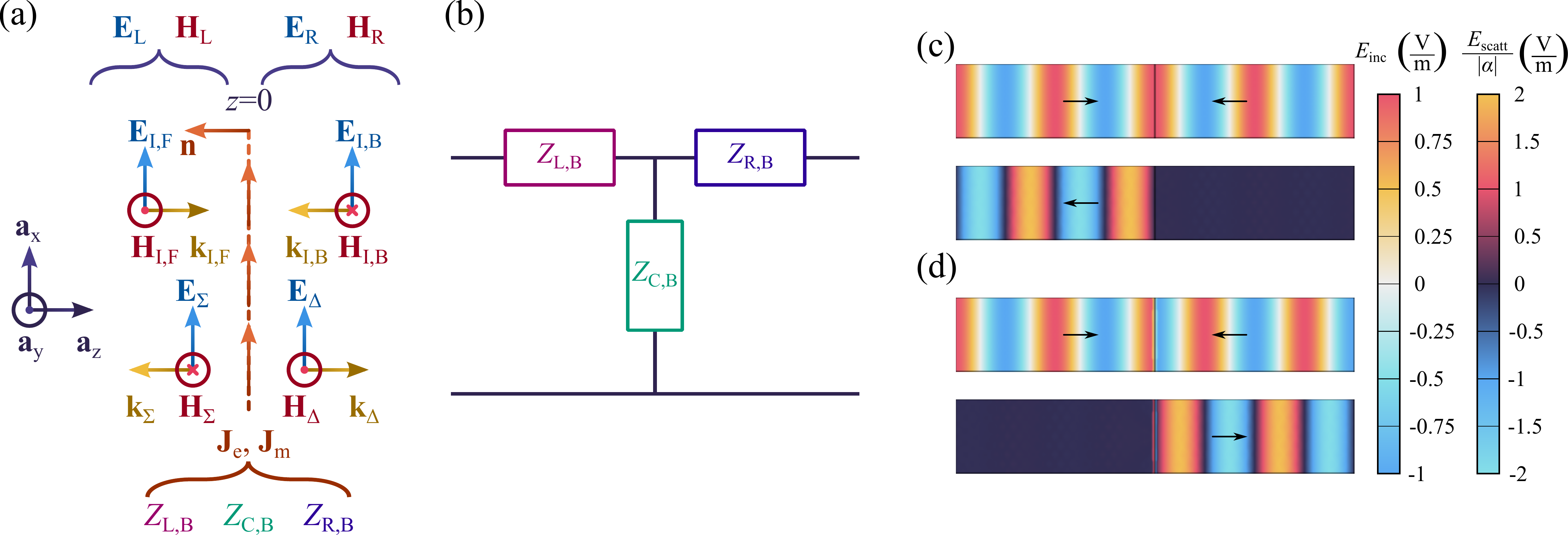}
  \caption{(a) A $\Delta\Sigma$ metasurface is a bianisotropic sheet capable of emulating a 180$^{\circ}$ hybrid junction functionalities under illumination by two plane waves. (b) The asymmetric reflection from such metasurface can be modeled as a two-port T-network, where the effects of excited electric and magnetic surface current densities $\_J\@{e}$ and $\_J\@{m}$ are governed by three equivalent surface impedances $Z\@{L,B}$, $Z\@{C,B}$,  and $Z\@{R,B}$ in a transmission-line model. (c) Simulation results obtained with COMSOL show that the device behaves as expected, producing total reflection when both incident waves illuminate the metasurface with the same phase and amplitude of the electric field, (d) while total forward scattering is produced with a $180^{\circ}$ phase difference between the sources. }
  \label{fig:biani_delta_sigma}
\end{figure*}

In this work, we discuss the concept of emulating bianisotropic response using an example of asymmetric reflection from a bianisotropic metasurface designed as a planar 180${^\circ}$ hybrid junction \cite{Pozar_2004} for plane waves. We show that it can be replicated using only metasurfaces with electric response illuminated by two coherent sources.

\section{$\Delta\Sigma$ Bianisotropic Metasurface}

In microwave engineering, a 180${^\circ}$ hybrid junction is a four-port device that combines two input waves and outputs waves proportional to their sum and difference \cite{Pozar_2004}. 
For plane waves, a two-port equivalent of this concept can be depicted using asymmetric
 scattering, either in transmission or reflection.
Such device, visualized in Fig.~\ref{fig:biani_delta_sigma}(a), functions for two input waves  simultaneously illuminating a single metasurface (one propagating in the ``forward'' direction with amplitude $E\@{F}$ and the other traveling in the ``backward'' direction with amplitude $E\@{B}$). We write the fields of the two incident waves as
\begin{subequations}
\label{eq:inc_sources}
\begin{align}
\begin{matrix}
 \_E\@{I,F}=E\@{F} e^{-j k_0 z} \_a\@{x}, & \_H\@{I,F}=\dfrac{E\@{F}}{\eta_0} e^{-j k_0 z} \_a\@{y},
 \end{matrix}\\
\begin{matrix}
 \_E\@{I,B}=E\@{B} e^{j k_0 z} \_a\@{x}, & \_H\@{I,B}=-\dfrac{E\@{B}}{\eta_0} e^{j k_0 z} \_a\@{y},
  \end{matrix}
\end{align}
\end{subequations}
where $\eta_0$ is the characteristic impedance of the medium surrounding the metasurface, and $k_0$ is the wavenumber of propagating waves. For this work, we assume the time harmonic convention $\exp (+j\omega t)$. 

The desired scattering, produced by the metasurface under this illumination, is equivalent to the sum ``$\Sigma$'' and difference ``$\Delta$'' of the incident waves. We write the scattered  fields as
\begin{subequations}
\label{e2}
\begin{equation}
 \_E\@{\Sigma}=\_E\@{R,F}+\_E\@{T,B}=\alpha \left(E\@{F} + E\@{B} \right)e^{j k_0 z} \_a\@{x}, 
\end{equation}
\begin{equation}
 \_E\@{\Delta}=\_E\@{R,B}+\_E\@{T,F}=\alpha \left(E\@{F} - E\@{B} \right) e^{-j k_0 z} \_a\@{x}.
\end{equation}
\end{subequations}
The fields are proportional (through the non-zero complex scaling factor $\alpha$) to the sum and difference of the amplitudes of the two incident waves.
Such interaction between incident waves can be realized using a single bianisotropic sheet that can be designed to support asymmetric transmission, asymmetric reflection, or arbitrary scattering. The bianisotropic sheet can be designed by imposing the desired scattering separately for each single source. For single forward illumination the scattered waves read
\begin{subequations}
\label{eq:scattering_forward}
\begin{equation}
\begin{matrix}
  \_E\@{T,F}=E\@{T,F} e^{-j k_0 z} \_a\@{x}, & \_H\@{T,F}=\dfrac{E\@{T,F}}{\eta_0} e^{-j k_0 z} \_a\@{y},
  \end{matrix}
\end{equation}
\begin{equation}
\begin{matrix}
  \_E\@{R,F}=E\@{R,F} e^{j k_0 z} \_a\@{x}, & \_H\@{R,F}=-\dfrac{E\@{R,F}}{\eta_0} e^{j k_0 z} \_a\@{y},
    \end{matrix}  
\end{equation}
\begin{equation}
\begin{matrix}
 E\@{T,F} =\tau\@{F} E\@{F}, & E\@{R,F} =\Gamma\@{F} E\@{F},
  \end{matrix}
\end{equation}  
\end{subequations}
where $\tau\@{F}$ ($\Gamma\@{F}$) is the transmission (reflection) coefficient with respect to the forward illumination. In the case of single backward illumination we write 
\begin{subequations}
\label{eq:scattering_backward}
\begin{equation}
  \begin{matrix}
  \_E\@{T,B}=E\@{T,B} e^{j k_0 z} \_a\@{x}, & \_H\@{T,B}=-\dfrac{E\@{T,B}}{\eta_0} e^{j k_0 z} \_a\@{y},
    \end{matrix}  
\end{equation}
\begin{equation}
  \begin{matrix}
  \_E\@{R,B}=E\@{R,B} e^{-j k_0 z} \_a\@{x}, & \_H\@{R,B}=\dfrac{E\@{R,B}}{\eta_0} e^{-j k_0 z} \_a\@{y},
  \end{matrix}
\end{equation}
\begin{equation}
\begin{matrix}
 E\@{T,B} =\tau\@{B} E\@{B}, & E\@{R,B} =\Gamma\@{B} E\@{B},
  \end{matrix}
\end{equation}  
\end{subequations}
where, accordingly, $\tau\@{B}$ and $\Gamma\@{B}$ are the transmission and reflection coefficients at backward illumination. In the case of asymmetric transmission ($\tau\@{F}\neq \tau\@{B}$), the sheet should be non-reciprocal, requiring external magnetic bias, active, or non-linear components \cite{Pozar_2004}. On the other hand, asymmetric reflection ($\Gamma\@{F}\neq \Gamma\@{B}$) requires presence of bianisotropy: magnetoelectric coupling, where the induced currents depend on both incident electric and magnetic fields \cite{Serdyukov_2001_bianisotropic, 
Asadchy_2018_bianisotropic}. Reciprocal magnetoelectric effects are classified into chiral effects, measured by the symmetric part of the coupling dyadic \cite{Lindell_1994,Simovksi_2020} and omega-coupling effects measured by the antisymmetric part of the coupling dyadic
\cite{Serdyukov_2001_bianisotropic,Simovksi_2020}. In both cases, the existence of coupling requires layers of non-zero thickness with specific geometric asymmetries. Specifically for the thought application we require asymmetry of reflection of linearly polarized incident waves, which corresponds to the presence of omega coupling \cite{Asadchy_2018_bianisotropic}. 
For the desired combined scattering of Eqs.~\eqref{e2}, we require a combination of symmetric transmission $\tau\@{F}=\tau\@{B}=\alpha$ and antisymmetric reflection $\Gamma\@{F}=-\Gamma\@{B}=\alpha$.
Such bianisotropic metasurface, isotropic in the transverse plane, can be described using the generalized sheet impedance conditions \cite{Grbic_2014_bianisotropic,Asadchy_2018_bianisotropic,yang_rahmat-samii_2019}
\begin{subequations}
\label{eq:eq_curr_biani_model}
\begin{equation}
 \_n\times\left(\_E\@{R}-\_E\@{L}\right)=\_J\@{m}, 
\end{equation}
\begin{equation}
 \_J\@{m}=Z\@{mm}\dfrac{\_H\@{R}+\_H\@{L}}{2}+\gamma\@{em}\_n\times\left[\dfrac{\_E\@{R}+\_E\@{L}}{2}\right], 
\end{equation}
\begin{equation}
 \_n\times\left(\_H\@{R}-\_H\@{L}\right)=-\_J\@{e},
\end{equation}
\begin{equation}
 \_J\@{e}=Y\@{ee}\dfrac{\_E\@{R}+\_E\@{L}}{2}+\chi\@{me}\_n\times\left[\dfrac{\_H\@{R}+\_H\@{L}}{2}\right].
\end{equation}
\end{subequations}
Here, $\_J\@{e}$ and $\_J\@{m}$ are the electric and magnetic surface current densities induced at the sheet, $Y\@{ee}$ is the electric sheet admittance, $Z\@{mm}$ the magnetic sheet impedance, and $\gamma\@{em}$ and $\chi\@{me}$ are the magnetoelectric coupling parameters. Furthermore,  $\_E\@{R}$ ($\_H\@{R}$) is the net tangential electric (magnetic) field at the right side of the interface of Fig.~\ref{fig:biani_delta_sigma}(a), while $\_E\@{L}$ ($\_H\@{L}$) is the corresponding net tangential electric (magnetic) field at the left side.
 These boundary conditions relate surface-averaged fields at the two sides of the layer, and they can be used to calculate reflected and transmitted fields at distances that are large compared with the array period \cite{Tretyakov_2003,yang_rahmat-samii_2019}. 
Solving the boundary conditions for each incident wave, as done in Appendix A, gives us the complex amplitudes of transmitted and reflected waves:
\begin{subequations}
\label{eq:coupling_parameters_forward}
\begin{equation}
 \tau\@{F}=\alpha= \dfrac{\eta_0 \left[4- Y\@{ee} Z\@{mm} -\gamma\@{em} \left(\chi\@{me}-2\right) -2 \chi\@{me}\right]}{2 Z\@{mm} + \eta_0 \left(4 + Y\@{ee} Z\@{mm} + \gamma\@{em} \chi\@{me}\right)+2 \eta_0^2 Y\@{ee}} , 
\end{equation}
\begin{equation}
 \Gamma\@{F} = \alpha = \dfrac{2 \left[Z\@{mm} - \eta_0 \left( \gamma\@{em}+\chi\@{me}\right) - \eta_0^2 Y\@{ee}\right]}{2 Z\@{mm} + \eta_0 \left(4 + Y\@{ee} Z\@{mm} + \gamma\@{em} \chi\@{me}\right)+2 \eta_0^2 Y\@{ee}} , 
\end{equation}
\end{subequations}
\begin{subequations}
\label{eq:coupling_parameters_backward}
\begin{equation}
 \tau\@{B}=\alpha = \dfrac{\eta_0 \left[4- Y\@{ee} Z\@{mm} -\gamma\@{em} \left(\chi\@{me}+2\right) +2 \chi\@{me}\right]}{2 Z\@{mm} + \eta_0 \left(4 + Y\@{ee} Z\@{mm} + \gamma\@{em} \chi\@{me}\right)+2 \eta_0^2 Y\@{ee}} , 
\end{equation}
\begin{equation}
 \Gamma\@{B}=-\alpha = \dfrac{2 \left[Z\@{mm} + \eta_0 \left( \gamma\@{em}+\chi\@{me}\right) - \eta_0^2 Y\@{ee}\right]}{2 Z\@{mm} + \eta_0 \left(4 + Y\@{ee} Z\@{mm} + \gamma\@{em} \chi\@{me}\right)+2 \eta_0^2 Y\@{ee}}.
\end{equation}
\end{subequations}
The required properties of such Delta-Sigma ($\Delta\Sigma$) bianisotropic metasurfaces defined by Eqs.~\eqref{eq:coupling_parameters_forward}--\eqref{eq:coupling_parameters_backward} are achieved when the sheet parameters take the form 
\begin{subequations}
\label{eq:delta_sigma_coupling_parameters}
\begin{equation}
 Y\@{ee}=\dfrac{2-4\alpha^2}{\eta_0\left(1+2\alpha+2\alpha^2\right)}, 
\end{equation}
\begin{equation}
 Z\@{mm}=\dfrac{\eta_0\left(2-4\alpha^2\right)}{1+2\alpha+2\alpha^2}, 
\end{equation}
\begin{equation}
 \gamma\@{em}=\chi\@{me}=-\dfrac{4\alpha}{1+2\alpha+2\alpha^2}.
\end{equation}
\end{subequations}
Contemplating the above expressions, two important properties are observed. First, we see that $Z_{\rm{mm}}=\eta_0^2Y_{\rm{ee}}$ regardless of the value of $\alpha$. Second, since the electromagnetic and magnetoelectric coupling parameters are the same ($\gamma\@{em}=\chi\@{me}$), this bianisotropic metasurface is reciprocal, and it can be modelled as a two-port T-network of Fig.~\ref{fig:biani_delta_sigma}(b) (see Appendices B--D). Hence, the $\Delta\Sigma$ bianisotropic metasurface is equivalent to three cascaded sheets (Left, Center, and Right), with surface impedances 
\begin{subequations}
\label{eq:biani_t_impedaces}
\begin{equation}
 Z\@{L,B}=\dfrac{Y\@{ee}Z\@{mm}-2 \gamma\@{em}+\gamma\@{em}^2}{2 Y\@{ee}}=\eta_0 \dfrac{1+2\alpha^2}{1-2\alpha^2},
\end{equation}
\begin{equation}
 Z\@{C,B}=\dfrac{4-Y\@{ee}Z\@{mm}-\gamma\@{em}^2}{4 Y\@{ee}}=\eta_0 \dfrac{2\alpha}{1-2\alpha^2}, 
\end{equation}
\begin{equation}
 Z\@{R,B}=\dfrac{Y\@{ee}Z\@{mm} + 2 \gamma\@{em}+\gamma\@{em}^2}{2 Y\@{ee}}=\eta_0 \dfrac{1-4\alpha+2\alpha^2}{1-2\alpha^2}. 
\end{equation}
\end{subequations}
\begin{figure*}[t]
  \centering
  \includegraphics[width=0.9\linewidth]{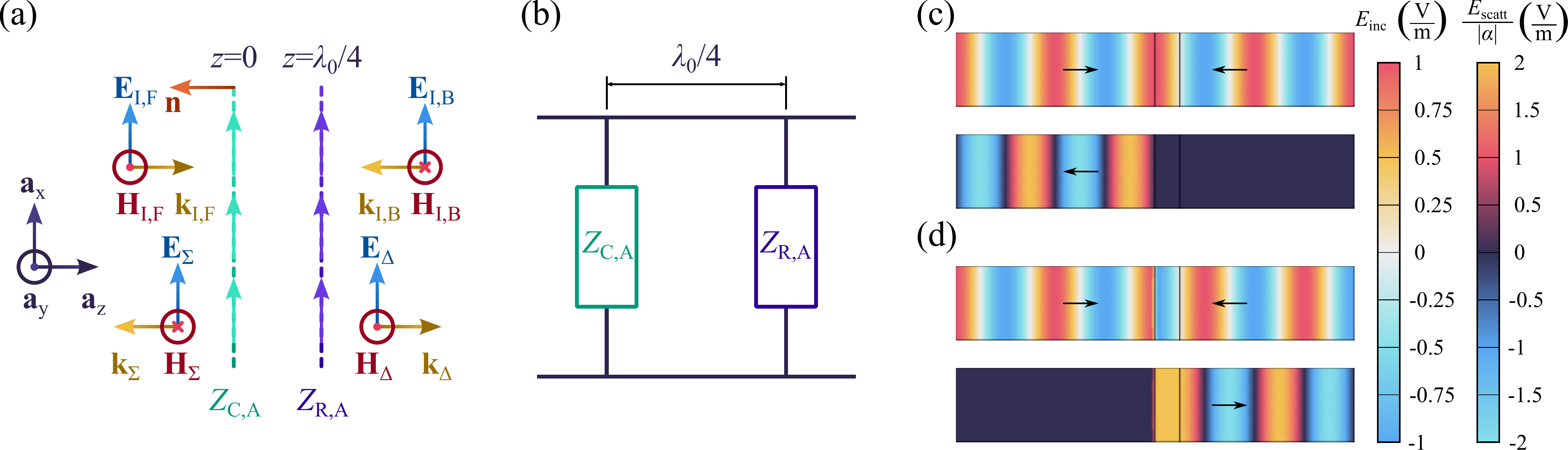}
  \caption{(a) A $\Delta\Sigma$ metasurface can be implemented as a combination of cascaded individual sheets with only electric current excitation. (b) In the case of a $\Delta\Sigma$ bianisotropic sheet of Eqs.~\eqref{eq:impedances_gamma_network}, one of the equivalent cascaded sheets can be suppressed, resulting in a metasurface pair separated $\lambda_0/4$ between them. (c)-(d) Simulation results show the same behaviour as of the $\Delta\Sigma$ metasurface pair implementation, using equivalent electric current densities with additional standing waves between the metasurfaces. For (c), using waves with the same amplitude and phase, only the sum scattering wave adds constructively, while the difference component fades. In (d), the additional $180^\circ$ phase difference results in a constructive interference in the difference scattering, as there is no sum component.}
  \label{fig:Paired_delta_sigma}
\end{figure*}
Two of the equivalent sheets have a magnetic response ($Z\@{L,B}$ and $Z\@{R,B}$) while the middle sheet has an electric response with sheet impedance $Z\@{C,B}$.
A lossless implementation of such interface requires the combined power of the output waves to be equal to the total input power (either with single or coherent forward-backward illumination). This condition is achieved when
\begin{equation}
  \alpha=\dfrac{1}{\sqrt{2}} e^{j\psi}, 
\end{equation}
where $\psi$ is an arbitrary phase angle. Based on this equation, the resulting equivalent impedances reduce to
\begin{subequations}
\begin{equation}
  Z\@{R,B}=\dfrac{j\eta_0\left(\cos \psi - \sqrt{2}\right)}{\sin \psi}, 
\end{equation}
\begin{equation}
 Z\@{C,B}=\dfrac{j\eta_0}{\sqrt{2} \sin \psi}, 
\end{equation}
\begin{equation}
 Z\@{L,B}=\dfrac{j\eta_0}{\tan \psi}.
\end{equation}
\end{subequations}
A simplified variant of such metasurface is achieved with $\alpha=\pm j/\sqrt{2}$ ($\psi=\pm \pi/2$), as in this case $Z\@{L,B}$ becomes zero, and the resulting $\Gamma$-model metasurface is fully characterized by the two remaining impedances 
\begin{equation}
\label{eq:impedances_gamma_network}
  \begin{matrix}
    Z\@{C,B} = \alpha \eta_0, & Z\@{R,B} = \dfrac{\eta_0}{\alpha}.
  \end{matrix}
\end{equation}

Simulation results presented in Figs.~\ref{fig:biani_delta_sigma}(c)-(d) show how the $\Delta\Sigma$ bianisotropic metasurface behaves under illumination by two incident waves with the same amplitudes with an arbitrary reference frequency $f_0=11.11$~GHz under two scenarios: when both waves have the same phase, and when there is a difference of $180^{\circ}$ phase between the  incidence sources. Both scenarios, simulated in COMSOL inside a parallel-plate waveguide consider an ideal $\Delta\Sigma$ bianisotropic metasurface by the sheet impedance boundary conditions~\eqref{eq:eq_curr_biani_model}. The amplitudes and phases for both scenarios were selected for their particular scattering outcome, as the scenario of Fig.~\ref{fig:biani_delta_sigma}(c) is expected to have only back-scattering ($E\@{\Delta}=0$), while the illumination corresponding to Fig.~\ref{fig:biani_delta_sigma}(d) should produce only forward scattering ($E\@{\Sigma}=0$).

\section{Asymmetric Scattering From Symmetric Structures}\label{section:pair}

The $\Delta\Sigma$ bianisotropic metasurface shown in Fig.~\ref{fig:biani_delta_sigma}(a) requires bianisotropic omega-coupling response defined in Eq.~\eqref{eq:eq_curr_biani_model}. In practice, this kind of response can be realized using $\Omega$-shaped particles \cite{Radi_2014_tayloring,Chen_2020_omega} or as arrays of cascaded planar meta-atoms \cite{Yazdi_2015_bianisotropic,Epstein_2016_arbitrary_omega,Popov_2019_omega,Xu_2020_discrete,Tamoor_2021_multifunctional}. While the $\Omega$-particle topology offers an intuitive understanding of its interaction with the electromagnetic fields, the latter topology is preferred due to its ease of fabrication. The $\Delta\Sigma$ bianisotropic metasurface of Eqs.~\eqref{eq:impedances_gamma_network} can be converted into a set of two independent cascaded sheets by applying the quarter-wave transformer to the magnetic component $Z\@{R,B}$ \cite{Pozar_2004}.
As a result, this $\Delta\Sigma$ metasurface pair can be realized as shown in Fig.~\ref{fig:Paired_delta_sigma}(a), where two sheets with identical surface impedances 
\begin{equation}
\begin{matrix}
  Z\@{C,A} = Z\@{C,B} = \alpha \eta_0, & Z\@{R,A} = \dfrac{\eta_0^2}{Z\@{R,B}} = \alpha \eta_0
 \end{matrix}  
\end{equation}
are separated by distance $d=\lambda_0/4$, where $\lambda_0$ is the operational wavelength.

\begin{figure*}[t]
  \centering
  \includegraphics[width=0.9\linewidth]{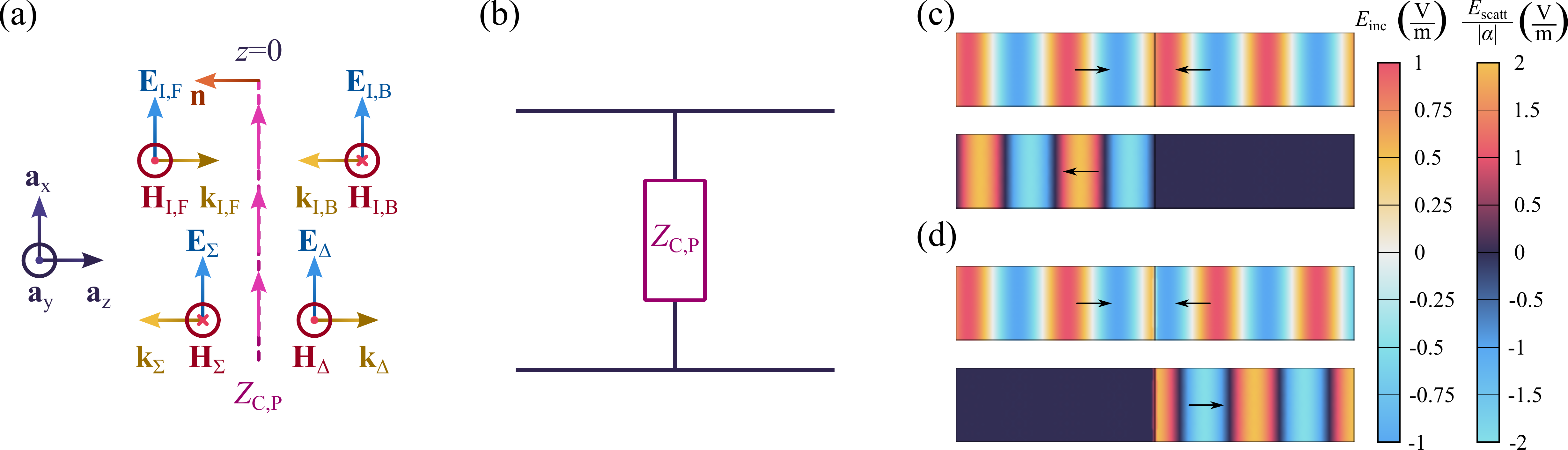}
  \caption{(a) By adding an additional phase-shift between the incident waves, it is possible to design a $\Delta\Sigma$ metasurface that exploits coherent illumination. (b) The equivalent transmission-line model for such sheet is a shunt impedance, which produces  symmetric reflection. (c) With the additional phase, a single non-bianisotropic metasurface is able to produce the desired asymmetric scattering in terms of producing back scattering when the incident waves are coherently in phase. (d) Likewise, total forward scattering is achieved using the same amplitudes but the incident waves have the phase difference of $180^{\circ}+2\phi$.}
  \label{fig:coherent_delta_sigma}
\end{figure*}

Interestingly, this metasurface pair is geometrically symmetric, as portrayed in Fig.~\ref{fig:Paired_delta_sigma}(b):  both parallel sheets have the same electric properties. It appears counter-intuitive that a symmetric device can produce asymmetric scattering. However, this ``pseudo-bianisotropy'' is possible because the considerable distance between the metasurfaces allows the use of coherent illumination with an asymmetrically-defined reference plane. In the absolute terms, phases of all input and output waves are counted from the same reference plane: the metasurface located at $z=0$ in Fig.~\ref{fig:Paired_delta_sigma}(a). Thus, from the point of view of each individual source, the forward wave (propagating in the $+z$ direction) has its reference plane at the first metasurface that the wave illuminates. By solving the boundary conditions at each sheet for single forward and backward illuminations (see Appendix E), it can be found that the amplitudes of the scattered waves of Eq.~\eqref{eq:scattering_forward} produced by the forward incident wave read
\begin{subequations}
\begin{equation}
  E\@{T,F}=E\@{F} \dfrac{2 Z\@{C,A}^2}{\eta_0^2+2 \eta_0 Z\@{C,A} + 2 Z\@{C,A}^2},
\end{equation}
\begin{equation}
  E\@{R,F}=- E\@{F} \dfrac{\eta_0^2}{\eta_0^2+2 \eta_0 Z\@{C,A} + 2 Z\@{C,A}^2}.
\end{equation}
\end{subequations}
However, the reference plane for the backward wave (propagating in the $-z$ direction) is located at the last metasurface that the wave reaches. In that case, the amplitudes of reflected and transmitted waves, described by Eqs.~\eqref{eq:scattering_backward}, read
\begin{subequations}
\label{eq:pair_forward_outer_fields}
\begin{equation}
  E\@{T,B}=E\@{B} \dfrac{2 Z\@{C,A}^2}{\eta_0^2+2 \eta_0 Z\@{C,A} + 2 Z\@{C,A}^2},
\end{equation}
\begin{equation}
  E\@{R,B}=- E\@{B} e^{j \phi\@{R,B}} \dfrac{\eta_0^2}{\eta_0^2+2 \eta_0 Z\@{C,A} + 2 Z\@{C,A}^2}.
\end{equation}
\end{subequations}
It can be noticed that the reflected wave $E\@{R,B}$ is defined with an additional phase of $\phi\@{R,B}=2 k_0 d= \pi$ with respect to the reflected wave produced under forward illumination. The resulting combined fields read
\begin{subequations}
\begin{equation}
\begin{split}
 \_E\@{\Sigma}&= \dfrac{- E\@{F}\eta_0^2+E\@{B} 2 Z\@{C,A}^2}{\eta_0^2+2 \eta_0 Z\@{C,A} + 2 Z\@{C,A}^2}e^{j k_0 z} \_a\@{x}\\
 &=\alpha\left(E\@{F} + E\@{B} \right)e^{j k_0 z} \_a\@{x},
\end{split} 
\end{equation}
\begin{equation}
\begin{split}
 \_E\@{\Delta}&= \dfrac{2 E\@{F} Z\@{C,A}^2- E\@{B} e^{j \phi\@{R,B}}\eta_0^2}{\eta_0^2+2 \eta_0 Z\@{C,A} + 2 Z\@{C,A}^2} e^{-j k_0 z} \_a\@{x}\\
 &=\alpha\left(E\@{F} - E\@{B} \right) e^{-j k_0 z} \_a\@{x}.
\end{split} 
\end{equation}
\label{eq:combined_scattering_pair}
\end{subequations}

Similarly as has been done for the $\Delta\Sigma$ bianisotropic metasurface, computations of induced surface current densities using COMSOL software show that the metasurface pair is capable of producing only back-scattering when it is illuminated with two incident waves with the same amplitude and phase, as presented in Fig.~\ref{fig:Paired_delta_sigma}(c). Total forward scattering is obtained after introducing a phase difference of 180$^{\circ}$ between the illuminating sources, as is done in Fig.~\ref{fig:Paired_delta_sigma}(d). 

\section{$\Delta\Sigma$ coherent metasurface}

Alternatively, a two-port analogy of an 180${^\circ}$ hybrid junction that creates scattering proportional to the sum and difference of the incident waves can be realized by a single non-bianisotropic sheet by external tuning of these incident waves. 
The device of Fig.~\ref{fig:coherent_delta_sigma}(a), namely a $\Delta\Sigma$ coherent metasurface, is designed as a thin sheet with electric surface impedance [defined by $Y\@{ee}=1/Z\@{C,P}$ and $Z\@{mm}=\gamma\@{em}=\chi\@{me}=0$ in Eq.~\eqref{eq:eq_curr_biani_model}] that is illuminated by two coherent sources with a matching phase $\phi$ in the form
\begin{subequations}
\label{eq:phased_sources}
\begin{equation}
 \_E\@{I,F}=E\@{F} e^{j \phi} e^{-j k_0 z} \_a\@{x}, 
\end{equation}
\begin{equation}
 \_E\@{I,B}=E\@{B} e^{-j \phi} e^{j k_0 z} \_a\@{x}. 
\end{equation}
\end{subequations}
The scatterings produced by each individual incidence source are combined using a relaxed definition of summation and subtraction of fields:
\begin{subequations}
\label{eq:relaxed_combined_scattering}
\begin{equation}
 \_E\@{\Sigma}=\alpha\@{\Sigma} \left(E\@{F} + E\@{B} \right)e^{j k_0 z} \_a\@{x}, 
\end{equation}
\begin{equation}
 \_E\@{\Delta}=\alpha\@{\Delta} \left(E\@{F} - E\@{B} \right) e^{-j k_0 z} \_a\@{x},
\end{equation}
\end{subequations}
where the sum and difference waves can have arbitrary amplitudes $\alpha\@{\Sigma}$ and $\alpha\@{\Delta}$, respectively. The resulting boundary conditions of Eq.~\eqref{eq:eq_curr_biani_model} can be split for each illumination source, so that the values of $Z\@{C,P}$, $\phi$, $\alpha\@{\Sigma}$ and $\alpha\@{\Delta}$ do not depend on $E\@{F}$ and $E\@{B}$. For single forward illumination scenario, the boundary conditions read 
\begin{subequations}
\label{eq:eigensolution_forward}
\begin{equation}
  e^{j\phi}+\alpha\@{\Sigma}=\alpha\@{\Delta},
\end{equation}
\begin{equation}
  2 Z\@{C,P}\left(e^{j \phi} -\alpha\@{\Sigma} -\alpha\@{\Delta}\right)=\eta_0\left(e^{j \phi} +\alpha\@{\Sigma} +\alpha\@{\Delta}\right),
\end{equation}
\end{subequations}
while for the backward incident wave we get
\begin{subequations}
\label{eq:eigensolution_backward}
\begin{equation}
  \alpha\@{\Sigma}=e^{-j\phi}-\alpha\@{\Delta},
\end{equation}
\begin{equation}
  2 Z\@{C,P}\left(e^{-j \phi} -\alpha\@{\Sigma} +\alpha\@{\Delta}\right)=\eta_0\left(e^{-j \phi} +\alpha\@{\Sigma} -\alpha\@{\Delta}\right).
\end{equation}
\end{subequations}
The boundary conditions of Eqs.~\eqref{eq:eigensolution_forward}--\eqref{eq:eigensolution_backward} are satisfied for combinations of sheet impedance and matching phase of the form
\begin{subequations}
\label{eq:coherent_conditions}
\begin{equation}
  \phi[n]=\dfrac{\pi\left(1+2n\right)}{4},\label{eq:coherent_conditions_phase}
\end{equation}
\begin{equation}
  Z\@{C,P}[n]=-\dfrac{j\eta_0 }{2 \tan \left(\phi[n]\right)},\label{eq:coherent_conditions_impedance}
\end{equation}
\end{subequations}
where $n$ is an integer. The corresponding values of $\alpha\@{\Sigma}$ and $\alpha\@{\Delta}$ read
\begin{subequations}
\label{eq:coherent_eigenequations}
\begin{equation}
  \alpha\@{\Sigma}=-j\sin \left(\phi[n]\right),
\end{equation}
\begin{equation}
  \alpha\@{\Delta}=\cos \left(\phi[n]\right).
\end{equation}
\end{subequations}

Unlike the previous structures, the $\Delta\Sigma$ functionality requires that both incident waves have an additional phase shift between them, defined by Eq.~\eqref{eq:coherent_conditions_phase}, as a form of ``coherent bianisotropy''. This single-sheet coherently-illuminated device was also tested in COMSOL with a similar setup, using a homogenized model of an electrically polarizable sheet. The difference scattering component is canceled when both incident waves have the same amplitude and a phase difference of $2\phi[n]$, as represented in Fig.~\ref{fig:coherent_delta_sigma}(c); while the sum component vanishes when the phase difference is equal to $2\phi[n]+180^{\circ}$, as shown in Fig.~\ref{fig:coherent_delta_sigma}(d). Please notice that, unlike the previous structures, the phases of these scattered waves are different, corresponding to the conditions given in Eqs.~\eqref{eq:coherent_eigenequations}.

\section{Stability under non-ideal illuminations}

\begin{figure}[t]
    \centering
    \includegraphics[width=0.8\linewidth]{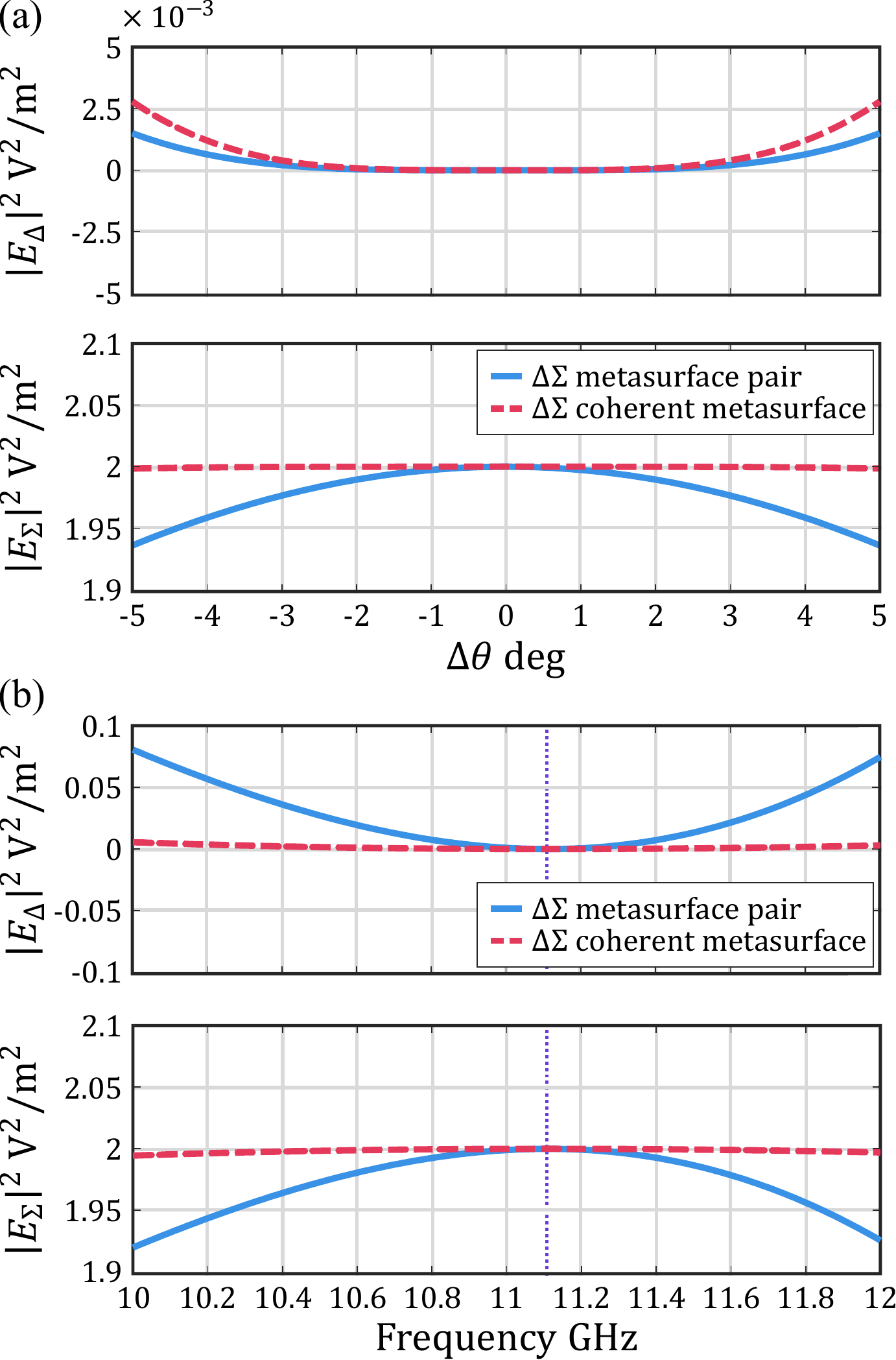}
    \caption{Stability of the $\Delta\Sigma$ metasurface pair and the $\Delta\Sigma$ coherent metasurface with respect to (a) shifts of the angle of incidence and (b) frequency deviations of the incidence waves. In both cases, it was assumed that the two incident waves have the same incidence angle or frequency. Both structures were designed for the  normal incidence ($\theta_i=0$) and the operational frequency of $f= 11.11$~GHz (dotted line), using frequency-dispersive impedance sheets.}
    \label{fig:angle_frequency_stability}
\end{figure}

Because the device operation requires coherent excitation, the performance of the metasurface pair and the coherent sheet degrades under non-ideal illuminations. Phase misalignment would happen when one of the sources has an additional phase $\Delta\Phi=\Phi-\Phi\@{ideal}$, introduced when the metasurface pair is not properly aligned to the reference plane, or, for the single-sheet device, the two incident waves do not have the proper phase difference. Under the assumption that only one of the incident waves is not properly matched (forward wave in this case), the phase difference is transferred into the complex amplitude of the wave $E'\@{F}=E\@{F}\exp{j \Delta\Phi}$. Therefore, the sum and difference responses, as described in Eqs.~\eqref{eq:relaxed_combined_scattering}, depend on this additional phase variation in the source. As a result, the amplitude of the scattered waves degrades from their expected values as $\Delta\Phi$ increases. Nevertheless, it can be found that this degradation in performance becomes noticeable only when $\vert \Delta\Phi \vert > 25^{\circ}$, where the amplitude of the sum component (using waves with the same amplitude and same original phase) falls below 95\% of the ideal value (see Appendix F). 

\begin{figure*}[t]
  \centering
  \includegraphics[width=1\linewidth]{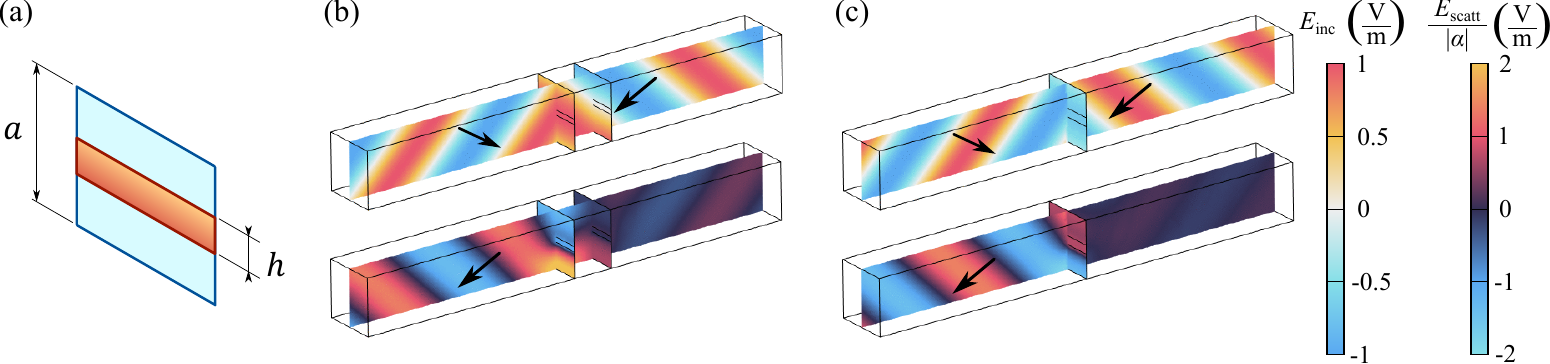}
  \caption{(a) The electric sheets can be modelled using anisotropic meta-atoms consisting of a metallic stripe with width $h$ and periodicity $a$. (b) The $\Delta\Sigma$ metasurface pair can be designed to operate for oblique incidence for TE-polarized waves, by adjusting the distance between the sheets. (c) With the respective considerations for the sheet impedance, the $\Delta\Sigma$ coherent sheet can also be designed to operate for oblique incidence. In both scenarios, we assumed two incident waves with $E\@{inc}=1$~V/m and zero phase difference at the reference plane, expecting  to obtain complete constructive interaction at the sum scattering, while the difference component  scatters  little-to-none power.}
  \label{fig:oblique_proof}
\end{figure*}

This performance degradation is also noticeable in scenarios where the frequency or the incident angle is not ideal. For such purpose, both coherently-illuminated structures were simulated in COMSOL as thin sheets with effective surface impedances, using two plane waves with the same amplitude, phase, frequency and, angle of incidence. In terms of frequency, as shown in Fig.~\ref{fig:angle_frequency_stability}(a), worse performance of the metasurface pair is due to the quarter-wave transformer implementation, where the use of two sheets makes the overall device frequency-resonant (the resonance is at 11.11~GHz in this example). As the impedances for each sheet should be designed for a particular angle of incidence, it is expected that the performance of both $\Delta\Sigma$ structures deteriorates  when the angle of incidence differs from the design one (in this case, both structures were designed for normal incidence). In the case of the metasurface pair, the weaker angular performance of Fig.~\ref{fig:angle_frequency_stability}(b) is also due to the quarter-wave transformer, as the standing wave in between the sheets depends on reflections produced by each single sheet. Nevertheless, both structures offer a broad frequency and angular performance range, as the performance of the metasurface pair is at least 95\% of the ideal level for incident angle variations up to  $10^\circ$ and  frequency variations up to about 1~GHz (approximately 9\% bandwidth). 

\section{Implementation for oblique angle of incidence}

Both coherent structures can be implemented using metasurfaces with only electric response. Additionally, they can be designed to operate for oblique illumination, mimicking a four-port device through specular reflection. For that purpose, example metasurfaces are designed to operate under transverse electric (TE) illumination at the angle of incidence $\theta\@{i}=45^{\circ}$, with the characteristic impedance $\eta=\eta_0/\cos \theta\@{i}$ and propagation constant $\beta=k_0\cos \theta\@{i}$ at the operational frequency of $f_0=10$~GHz \cite{Tretyakov_2003}. In accordance to the change in the propagation constant, the distance between metasurfaces for the $\Delta\Sigma$ metasurface pair is increased to $d=\lambda_0/4\cos\theta\@{i}$.

In this setup, each sheet can be realized using metallic stripes of Fig.~\ref{fig:oblique_proof}(a), with a periodicity of $a=0.5 \lambda_0$ (14.99~mm) for all the required metasurfaces. This meta-atom is designed based on the homogenization principle, where the array of meta-atoms can be replaced by a uniform sheet characterized by an effective surface impedance. The homogenized layer produces the same fields as the actual array of meta-atoms at distances that are large as compared with the array period \cite{Tretyakov_2003,Cuesta_2020_non_scattering_cavities}. The width of the metallic stripe was tuned to give the desired scattering under oblique incidence. For the metasurface pair, the stripe width was found to be $h= 0.045 \lambda_0$ (1.35~mm), while for the single coherent sheet the width was $h= 0.076 \lambda_0$ (2.28~mm). The simulation results, realized in COMSOL using Perfect Electric Conductor (PEC) stripes, corroborate the expected performance for TE illuminating plane waves, whose electric field is parallel to the metal strips. In the case of the metasurface pair of Fig.~\ref{fig:oblique_proof}(b), illuminated by two incident waves with the same amplitude and phase, the sum component picks up 94\% of the incident power. Even better performance was achieved for the single coherent sheet of Fig.~\ref{fig:oblique_proof}(c), which under similar illumination (with an added phase-shift of $\phi=5\pi/4$), creates the sum component that carries 99\% of the total power.


\section{Coherent illumination emulating chirality effects}

\begin{figure*}[t]
  \centering
  \includegraphics[width=1\linewidth]{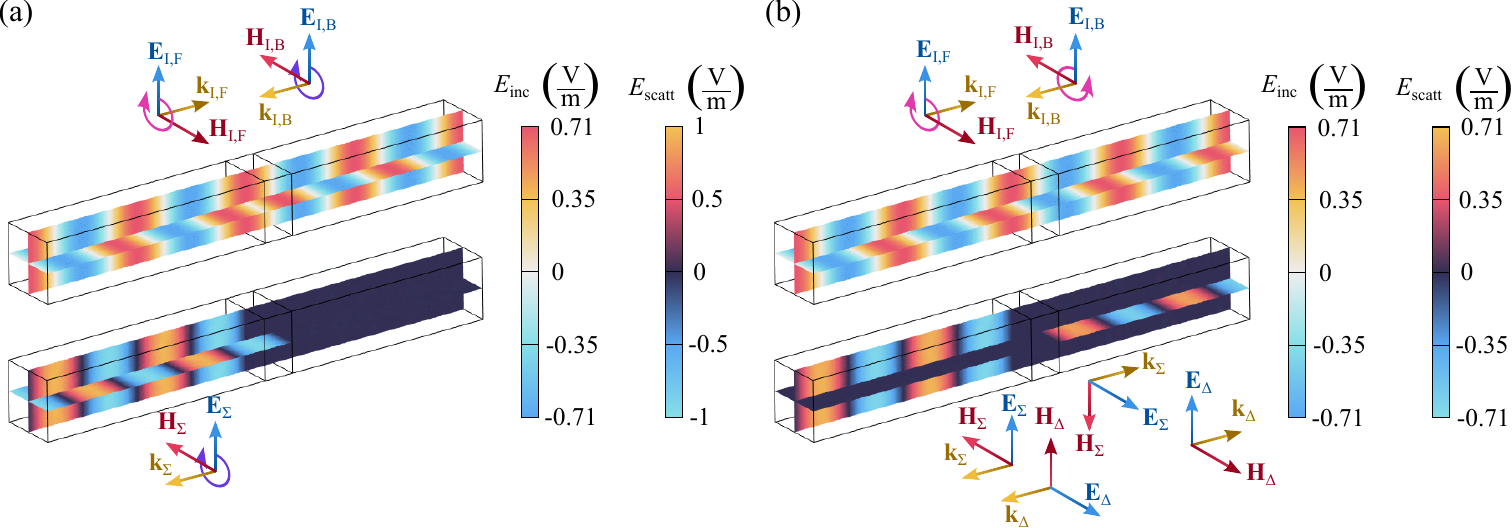}
  \caption{ (a) The $\Delta\Sigma$ metasurface pair can also emulate some chiral effects under coherent illumination. Using isotropic sheets, it is capable to combine constructively a left-handed wave with a right-handed one when their phases are the same at the first sheet. In this scenario, the concept was proved numerically via COMSOL using the $\Delta\Sigma$ metasurface pair of Section~\ref{section:pair} as two homogenized sheets carrying electric surface current. (b) Additionally, if the metasurface pair is illuminated by two circularly-polarized waves with the same handedness (right-hand polarization in this case), the resulting scattered waves have a vertical or a horizontal polarization depending on the side of the device and the phase difference between the illuminating waves.}
  \label{fig:circular_proof}
\end{figure*}

The $\Delta\Sigma$ structure explained in the previous sections was introduced as an example of the use of coherent illumination to emulate bianisotropic coupling of the omega type. In that case, linearly polarized incident waves reflect differently, depending on which side of the metasurface they illuminate. Next,  we show that this geometrically symmetric metasurface can emulate also effects of chiral bianisotropic coupling, with the use of coherent illumination by a circularly polarized control wave.  

In the case of an incident wave with an arbitrary polarization, it is found that the reflected wave has the opposite handedness of the incident wave, with transmission and reflection coefficients following the expressions of Eqs.~\eqref{eq:coupling_parameters_forward}--\eqref{eq:coupling_parameters_backward} (see Appendix G). Using these equations, we see that if the layer is coherently illuminated by two circularly-polarized waves with the  opposite handedness (one is right-handed and the other one is left-handed), the scattered waves at each side of the interface have the same handedness and they can merge together. Therefore, for the $\Delta\Sigma$ metasurface under coherent illumination by waves of the same phase and amplitude, forward scattering vanishes, and only backward scattering remains, which has the same handedness as the backward incident wave. In Fig.~\ref{fig:circular_proof}(a), we confirm this result in numerical simulations, using a similar setup of the ideal $\Delta\Sigma$ metasurface pair of Section~\ref{section:pair}. As seen, the numerical results are fully in agreement with the theoretical expectations. Note that this control over handedness of scattering is enabled by the use of a control wave of a certain handedness, since the sheets are uniform, isotropic, and symmetric with respect to mirror inversion. Without coherent illumination by a handed control wave, no chirality effects are possible.   

As another interesting example, we show that it is also possible to spatially decompose a circularly polarized wave into two orthogonal linearly polarized scattered waves, again using coherent illumination of a non-chiral layer. To realize such effect, the two illuminating coherent waves should have the same handedness. Due to this property and the inversion of the handedness in reflection, we see that the $\Delta\Sigma$ layer  splits the incident circularly polarized wave into two linearly polarized waves, propagating into the opposite directions, 
as illustrated in Fig.~\ref{fig:circular_proof}(b). This happens because the vertically-polarized  difference wave propagates  in the forward direction (resulting in vanishing of this wave) while the sum wave is in the backward direction. However, for the  horizontal polarization, the situation is opposite and the sum wave propagates in the forward direction, while the difference wave (with zero total amplitude) is in the backward direction. Therefore, the vertical and horizontal components are spatially separated. It is important to note that the symmetry of reflection of right- and left circularly polarized waves illuminating one side of a non-chiral layer is broken and controlled by the helicity of the coherent control wave that illuminates the other side. 
Similarly to the previous example, without the presence of coherent control wave such polarization splitting by an isotropic in the plane layer would require a mirror-asymmetric (chiral) structure. Thus, we see that coherent illumination emulates breaking of the mirror-reflection symmetry of metasurfaces. 


\section{Conclusions}

In summary, we have proposed a possibility to emulate and control bianisotropic response from non-bianisotropic objects using coherent illumination by two waves. This basically means that symmetry of scattering phenomena can be controlled by adjusting the phase or amplitude of the control wave. In the first part, we discussed how asymmetric reflection from bianisotropic metasurfaces can be emulated using coherent illumination. As a representative example, we considered a metasurface analogy of a 180$^{\circ}$ hybrid junction, called $\Delta\Sigma$ metasurface. The amplitudes of waves scattered by such a sheet into opposite directions are proportional to the sum and difference of the amplitudes of two illuminating waves. If we require that such device is a metasurface with electrically negligible thickness, this functionality requires bianisotropic properties (reciprocal omega coupling) in the metasurface. However, if we allow electrically considerable thickness, the equivalent response can be realized using two parallel metasurfaces that both maintain only electric surface currents. In this case, the required bianisotropic coupling is emulated by an additional phase shift of waves propagating between the two sheets. We call this effect \emph{pseudo-bianisotropy}. Furthermore, we have shown that if we illuminate a metasurface by two coherent plane waves with a particular value of the phase difference of the incident electric fields, the same functionality of an asymmetrically reflecting $\Delta\Sigma$ metasurface can be realized in a single non-bianisotropic, infinitely thin sheet. Thus, specific coherent illumination of a non-bianisotropic sheet by two coherent waves emulates bianisotropic properties of metasurfaces that are required for breaking symmetry of reflection properties of two sides of thin sheets. In addition, we have shown that some effects usually associated with chirality can also be emulated using coherent illumination. This concept of \emph{coherent bianisotropy} allows us to realize effects of bianisotropy in simpler non-bianisotropic metasurfaces with either electric or magnetic response. Since these effects exist only in the presence of a proper control wave, they are optically tunable. In this work, we also presented a proof-of-concept of coherent bianisotropy at oblique illumination of non-bianisotropic metasurfaces. Finally we note that although in this work we have considered planar layers, the introduced concept of coherent bianisotropy is general and can be applied to asymmetric scattering from any bianisotropic object, for example, from small bianisotropic particles.

\section{Acknowledgements}
This work was supported in part by the Academy of Finland under grant 330260 and by Nokia Foundation under scholarship 20200224.

\renewcommand\thefigure{A.\arabic{figure}}    
\setcounter{figure}{0}   

\renewcommand\theequation{A\arabic{equation}}    
\setcounter{equation}{0}

\section*{APPENDIX A: Coupling parameters for a bianisotropic sheet with arbitrary scattering}

\begin{figure}[h]
    \centering    \includegraphics[width=1\linewidth]{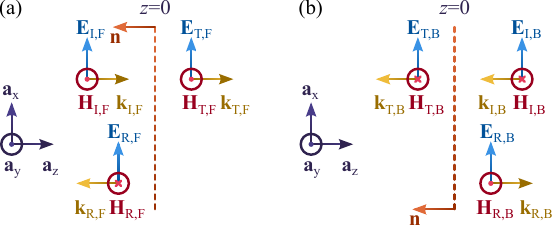}
    \caption{A bianisotropic sheet with linear response can be characterized in terms the scattering produced by each individual source. In the case of this work, it is assumed that the bianisotropic sheet is illuminated in the (a) forward and (b) backward direction.}
    \label{fig:sche_biani}
\end{figure}

The bianisotropic metasurface of Fig.~\ref{fig:sche_biani} is characterized using two independent incident waves, each one illuminating from opposite sides of the interface. The corresponding electric fields are as defined in Eqs.~\eqref{eq:inc_sources}.
%
%
The forward wave in Fig.~\ref{fig:sche_biani}(a) produces two scattering waves, one transmitting wave (with the same propagating direction as the incident wave) and a reflected wave (traveling in the opposite direction). The electric fields for both waves can be written as in Eqs.~\eqref{eq:scattering_forward}.
%
%
In the case of single forward illumination, the boundary conditions of Eqs.~\eqref{eq:eq_curr_biani_model} are reduced into the form 
\begin{subequations}
\label{eq:eq_bound_biani_forward}
\begin{equation}
\begin{split}
  -\left(E\@{T,F}-E\@{F}-E\@{R,F}\right)=&\dfrac{Z\@{mm}}{2\eta_0}\left(E\@{T,F}+E\@{F}-E\@{R,F}\right)\\
  &-\dfrac{\gamma\@{em}}{2}\left(E\@{T,F}+E\@{F}+E\@{R,F}\right),
  \end{split}
\end{equation}
\begin{equation}
\begin{split}
  \dfrac{1}{\eta_0}\left(E\@{T,F}-E\@{F}+E\@{R,F}\right)=&-\dfrac{Y\@{ee}}{2}\left(E\@{T,F}+E\@{F}+E\@{R,F}\right)\\
  &-\dfrac{\chi\@{me}}{2\eta_0}\left(E\@{T,F}+E\@{F}-E\@{R,F}\right).
  \end{split}
\end{equation}
\end{subequations}
From Eqs.~\eqref{eq:eq_bound_biani_forward} can be extracted a transmission coefficient $\tau\@{F}$ and a reflection coefficient $\Gamma\@{F}$, with respect to the forward-illuminating wave, which expressions are summarized in Eqs.~\eqref{eq:coupling_parameters_forward}.
%
%

On the other side, the scattering produced by an incident wave traveling in the backward direction, as portrayed in Fig.~\ref{fig:sche_biani}(b), can be described as in Eqs.~\eqref{eq:scattering_backward}.
%
%
The boundary conditions of Eqs.~\eqref{eq:eq_curr_biani_model} for the backward-illuminating case are reduced into
\begin{subequations}
\label{eq:eq_bound_biani_backward}
\begin{equation}
\begin{split}
  \left(E\@{B}+E\@{R,B}-E\@{T,B}\right)=&\dfrac{Z\@{mm}}{2\eta_0}\left(E\@{B}-E\@{R,B}+E\@{T,B}\right)\\
  &+\dfrac{\gamma\@{em}}{2}\left(E\@{B}+E\@{R,B}+E\@{T,B}\right),
  \end{split}
\end{equation}
\begin{equation}
\begin{split}
  \dfrac{1}{\eta_0}\left(E\@{B}-E\@{R,B}-E\@{T,B}\right)=&\dfrac{Y\@{ee}}{2}\left(E\@{B}+E\@{R,B}+E\@{T,B}\right)\\
  &-\dfrac{\chi\@{me}}{2\eta_0}\left(E\@{B}-E\@{R,B}+E\@{T,B}\right);
  \end{split}
\end{equation}
\end{subequations}
and are solved similarly in terms of transmission ($\tau\@{B}$) and reflection ($\Gamma\@{B}$) coefficients, described as in Eqs.~\eqref{eq:coupling_parameters_backward}.
%

Therefore, if the desired scattering is known for forward and backward direction, the coupling parameters of Eqs.~\eqref{eq:eq_curr_biani_model} can be reduced from Eqs.\eqref{eq:coupling_parameters_forward} and \eqref{eq:coupling_parameters_backward} into
\begin{subequations}
\label{eq:coupling_parameters_biani}
\begin{equation}
  Y\@{ee}=\dfrac{2}{\eta_0} \dfrac{(1-\Gamma\@{F})(1-\Gamma\@{B})-\tau\@{F}\tau\@{B}}{(1+\tau\@{F})(1+\tau\@{B})-\Gamma\@{F}\Gamma\@{B}},
\end{equation}
\begin{equation}
  Z\@{mm}=2 \eta_0 \dfrac{(1+\Gamma\@{F})(1+\Gamma\@{B})-\tau\@{F}\tau\@{B}}{(1+\tau\@{F})(1+\tau\@{B})-\Gamma\@{F}\Gamma\@{B}},
\end{equation}
\begin{equation}
  \gamma\@{em}=- 2 \dfrac{\Gamma\@{F}-\Gamma\@{B}-\tau\@{F}+\tau\@{B}}{(1+\tau\@{F})(1+\tau\@{B})-\Gamma\@{F}\Gamma\@{B}},
\end{equation}
\begin{equation}
  \chi\@{me}=- 2 \dfrac{\Gamma\@{F}-\Gamma\@{B}+\tau\@{F}-\tau\@{B}}{(1+\tau\@{F})(1+\tau\@{B})-\Gamma\@{F}\Gamma\@{B}}.
\end{equation}
\end{subequations}
In the case of a reciprocal sheet with asymmetric reflection ($\tau\@{F}=\tau\@{B}=\tau$), the coupling parameters take the form
\begin{subequations}
\label{eq:coupling_parameters_reciprocal}
\begin{equation}
  Y\@{ee}=\dfrac{2}{\eta_0} \dfrac{(1-\Gamma\@{F})(1-\Gamma\@{B})-\tau^2}{\left(1+\tau\right)^2-\Gamma\@{F}\Gamma\@{B}},
\end{equation}
\begin{equation}
  Z\@{mm}=2 \eta_0 \dfrac{(1+\Gamma\@{F})(1+\Gamma\@{B})-\tau^2}{\left(1+\tau\right)^2-\Gamma\@{F}\Gamma\@{B}},
\end{equation}
\begin{equation}
  \gamma\@{em}=\chi\@{me}=- 2 \dfrac{\Gamma\@{F}-\Gamma\@{B}}{\left(1+\tau\right)^2-\Gamma\@{F}\Gamma\@{B}}.
\end{equation}
\end{subequations}
Please notice that in this scenario we have $\gamma\@{em}=\chi\@{me}$, which allows to characterize the bianisotropic sheet as a two-port reciprocal device, using models based on T or $\Pi$ networks. It can be demonstrated that the expressions derived in Eqs.~\eqref{eq:biani_t_impedaces} are obtained when $\tau=\alpha$ and $\Gamma\@{B}=-\Gamma\@{F}=\alpha$. In the complementary case, where the bianisotropic sheet is non-reciprocal but it has symmetric reflection ($\Gamma\@{F}=\Gamma\@{B}=\Gamma$), the coupling parameters read
\begin{subequations}
\label{eq:coupling_parameters_nonreciprocal}
\begin{equation}
  Y\@{ee}=\dfrac{2}{\eta_0} \dfrac{\left(1-\Gamma\right)^2-\tau\@{F}\tau\@{B}}{(1+\tau\@{F})(1+\tau\@{B})-\Gamma^2},
\end{equation}
\begin{equation}
  Z\@{mm}=2 \eta_0 \dfrac{\left(1+\Gamma\right)^2-\tau\@{F}\tau\@{B}}{(1+\tau\@{F})(1+\tau\@{B})-\Gamma^2},
\end{equation}
\begin{equation}
  \gamma\@{em}=-\chi\@{me}= 2 \dfrac{\tau\@{F}-\tau\@{B}}{(1+\tau\@{F})(1+\tau\@{B})-\Gamma^2}.
\end{equation}
\end{subequations}
It is found that a symmetric reflection requires the condition $\gamma\@{em}=-\chi\@{me}$ instead.

\renewcommand\thefigure{B.\arabic{figure}}    
\setcounter{figure}{0}   

\renewcommand\theequation{B\arabic{equation}}    
\setcounter{equation}{0}

\section*{APPENDIX B: Equivalent boundary conditions for bianisotropic metasurfaces modelled as two-port T-networks}

Metasurfaces with bianisotropic response can be viewed as combinations of alternating sheets with electric or magnetic response. The distance between the sheets are assumed to be negligible and the sheets are effectively homogeneous.

\begin{figure}[h]
    \centering    \includegraphics[width=1\linewidth]{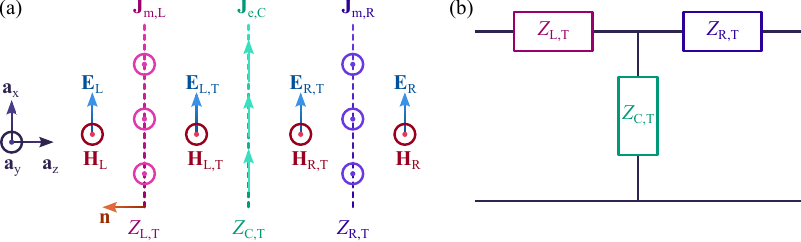}
    \caption{A general reciprocal T-network can be envisioned as three tightly cascaded current sheets. The shunt components model sheets with magnetic surface currents on them, while sheets with electric surface currents correspond to the behavior of the shunt component.}
    \label{fig:sche_t_ms}
\end{figure}

The configuration of Fig.~\ref{fig:sche_t_ms} contains three cascaded surface impedance sheets: two mangetically-polarizable and one electrically-polarizable placed between the former two. The boundary conditions for each impedance sheet read:
\begin{subequations}
\label{eq:bound_cond_lt}
\begin{equation}
  \_n\times\left(\_E\@{L,T}-\_E\@{L}\right)=\_J\@{m,L}, 
\end{equation}
\begin{equation}
  \_n\times\left(\_H\@{L,T}-\_H\@{L}\right)=0,
\end{equation}
\begin{equation}
  \_J\@{m,L}=Z\@{L,T}\dfrac{\_H\@{L,T}+\_H\@{L}}{2}=Z\@{L,T} \_H\@{L};
\end{equation}
\end{subequations}
\begin{subequations}
\label{eq:bound_cond_ct}
\begin{equation}
  \_n\times\left(\_E\@{R,T}-\_E\@{L,T}\right)=0, 
\end{equation}
\begin{equation}
  \_n\times\left(\_H\@{R,T}-\_H\@{L,T}\right)=-\_J\@{e,C},
\end{equation}
\begin{equation}
   \_J\@{e,C}=\dfrac{\_E\@{R,T}+\_E\@{L,T}}{2Z\@{C,T}};
\end{equation}
\end{subequations}
and
\begin{subequations}
\label{eq:bound_cond_rt}
\begin{equation}
  \_n\times\left(\_E\@{R}-\_E\@{R,T}\right)=\_J\@{m,R}, 
\end{equation}
\begin{equation}
  \_n\times\left(\_H\@{R}-\_H\@{R,T}\right)=0,
\end{equation}
\begin{equation}
  \_J\@{m,R}=Z\@{R,T}\dfrac{\_H\@{R}+\_H\@{R,T}}{2}=Z\@{R,T} \_H\@{R},
\end{equation}
\end{subequations}
where $\_E\@{L,T}$ ($\_H\@{L,T}$) is the net  tangential electric (magnetic) field between the left and center sheets; while $\_E\@{R,T}$ ($\_H\@{R,T}$) is the corresponding electric (magnetic) field between the center and right sheets. The tangential fields $\_E\@{L,T}$ and $\_H\@{L,T}$ are determined using the boundary conditions on the left sheet of Eqs.~\eqref{eq:bound_cond_lt}, written as
\begin{subequations}
\begin{equation}
  \_n\times\_H\@{L,T}=\_n\times\_H\@{L}, 
\end{equation}
\begin{equation}
  \_n\times\_E\@{L,T}=\_n\times\_E\@{L}+Z\@{L,T} \_H\@{L}.
\end{equation}
\end{subequations}
Similarly, tangential fields $\_E\@{R,T}$ and $\_H\@{R,T}$ are related through the right sheet's boundary conditions of Eqs.~\eqref{eq:bound_cond_rt} as
\begin{subequations}
\begin{equation}
  \_n\times\_H\@{R,T}=\_n\times\_H\@{R}, 
\end{equation}
\begin{equation}
 \_n\times\_E\@{R,T}=\_n\times\_E\@{R}-Z\@{R,T} \_H\@{R}.
\end{equation}
\end{subequations}
Therefore, equivalent boundary conditions for the whole setup can be established replacing the values of the inner tangential fields into the middle sheet boundary conditions of Eqs.~\eqref{eq:bound_cond_ct}, leading to
\begin{subequations}
\begin{equation}
  \_n\times\left(\_E\@{R}-\_E\@{L}\right)=Z\@{R,T} \_H\@{R} + Z\@{L,T} \_H\@{L}, 
\end{equation}
\begin{equation}
\begin{split}
  \_n\times\left(\_H\@{R}-\_H\@{L}\right)=-\dfrac{1}{2Z\@{C,T}}&\left[\_E\@{R}+\_E\@{L}\right.\\
  &\left.+\_n\times\left(Z\@{R,T} \_H\@{R}-Z\@{L,T} \_H\@{L}\right)\right].
  \end{split}
\end{equation}
\end{subequations}
%

The boundary conditions can be rearranged so that the surface current densities can be defined in terms of Eqs.~\eqref{eq:eq_curr_biani_model},
%
%
in the case of the T-network these parameters read
\begin{subequations}
\label{eq:rho_t_model}
\begin{equation}
  Y\@{ee,T}=\dfrac{4}{Z\@{L,T}+ 4 Z\@{C,T} + Z\@{R,T} }, 
\end{equation}
\begin{equation}
  \chi\@{me,T}=\gamma\@{em,T}=\dfrac{2\left(Z\@{R,T}-Z\@{L,T}\right)}{Z\@{L,T}+ 4 Z\@{C,T} + Z\@{R,T}},
\end{equation}
\begin{equation}
  Z\@{mm,T}=\dfrac{4\left(Z\@{L,T}Z\@{C,T}+Z\@{C,T}Z\@{R,T}+Z\@{R,T}Z\@{L,T}\right)}{Z\@{L,T}+ 4 Z\@{C,T} + Z\@{R,T}}.
\end{equation}
\end{subequations}
Notice that the relation $\chi\@{me,T}=\gamma\@{em,T}$ is valid due the reciprocal nature of the T-network. Complementarily, the network impedances can be described in terms of the coupling parameters as
\begin{subequations}
\label{eq:t_model_eq_impedances}
\begin{equation}
  Z\@{L,T}=\dfrac{Y\@{ee,T}Z\@{mm,T}-2 \gamma\@{em,T}+\gamma\@{em,T}^2}{2 Y\@{ee,T}}, 
\end{equation}
\begin{equation}
  Z\@{C,T}=\dfrac{4-Y\@{ee,T}Z\@{mm,T}-\gamma\@{em,T}^2}{4 Y\@{ee,T}},
\end{equation}
\begin{equation}
  Z\@{R,T}=\dfrac{Y\@{ee,T}Z\@{mm,T} + 2 \gamma\@{em,T}+\gamma\@{em,T}^2}{2 Y\@{ee,T}}.
\end{equation}
\end{subequations}
%

\renewcommand\thefigure{C.\arabic{figure}}    
\setcounter{figure}{0}   

\renewcommand\theequation{C\arabic{equation}}    
\setcounter{equation}{0}

\section*{APPENDIX C: Equivalent boundary conditions for bianisotropic metasurfaces modelled as two-port $\Pi$-networks}

\begin{figure}[!h]
    \centering    \includegraphics[width=1\linewidth]{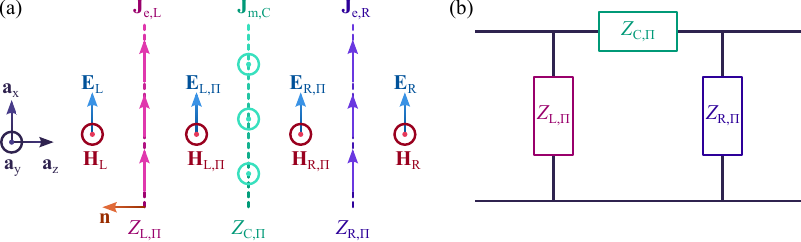}
    \caption{A $\Pi$-network is equivalent to three cascaded current sheets: two with electric currents and one with magnetic currents, positioned between the electric ones. The distances between sheets are negligible, so that there are no phase shifts between insertions.}
    \label{fig:sche_pi_ms}
\end{figure}

On the other hand, the configuration of Fig.~\ref{fig:sche_pi_ms} is formed by three cascaded surface impedances: two electrically-polarizable sheets and one magnetically-polarizable sheet placed between the former two. The boundary conditions for each sheet read
\begin{subequations}
\label{eq:bound_cond_lpi}
\begin{equation}
  \_n\times\left(\_E\@{L,\Pi}-\_E\@{L}\right)=0, 
\end{equation}
\begin{equation}
  \_n\times\left(\_H\@{L,\Pi}-\_H\@{L}\right)=-\_J\@{e,L},
\end{equation}
\begin{equation}
  \_J\@{e,L}=\dfrac{\_E\@{L,\Pi}+\_E\@{L}}{2Z\@{L,\Pi}}=\dfrac{\_E\@{L}}{Z\@{L,\Pi}};
\end{equation}
\end{subequations}
\begin{subequations}
\label{eq:bound_cond_cpi}
\begin{equation}
  \_n\times\left(\_E\@{R,\Pi}-\_E\@{L,\Pi}\right)=\_J\@{m,C}, 
\end{equation}
\begin{equation}
  \_n\times\left(\_H\@{R,\Pi}-\_H\@{L,\Pi}\right)=0,
\end{equation}
\begin{equation}
  \_J\@{m,C}=Z\@{C,\Pi}\dfrac{\_H\@{R,\Pi}+\_H\@{L,\Pi}}{2};
\end{equation}
\end{subequations}
and
\begin{subequations}
\label{eq:bound_cond_rpi}
\begin{equation}
  \_n\times\left(\_E\@{R}-\_E\@{R,\Pi}\right)=0, 
\end{equation}
\begin{equation}
  \_n\times\left(\_H\@{R}-\_H\@{R,\Pi}\right)=-\_J\@{e,R},
\end{equation}
\begin{equation}
  \_J\@{e,R}=\dfrac{\_E\@{R}+\_E\@{R,\Pi}}{2Z\@{R,\Pi}}=\dfrac{\_E\@{R}}{Z\@{R,\Pi}}.
\end{equation}
\end{subequations}
Similarly to  the above derivations, the inner tangential fields $\_E\@{L,\Pi}$ and $\_H\@{L,\Pi}$ are determined through the left sheet's boundary conditions of Eqs.~\eqref{eq:bound_cond_lpi}, written as
\begin{subequations}
\begin{equation}
  \_n\times\_E\@{L,\Pi}=\_n\times\_E\@{L}, 
\end{equation}
\begin{equation}
  \_n\times\_H\@{L,\Pi}=\_n\times\_H\@{L}-\dfrac{\_E\@{L}}{Z\@{L,\Pi}}.
\end{equation}
\end{subequations}
Accordingly, tangential fields $\_E\@{R,\Pi}$ and $\_H\@{R,\Pi}$ are related through the boundary conditions of the right sheet, based on Eqs.~\eqref{eq:bound_cond_rpi}, as
\begin{subequations}
\begin{equation}
  \_n\times\_E\@{R,\Pi}=\_n\times\_E\@{R}, 
\end{equation}
\begin{equation}
  \_n\times\_H\@{R,\Pi}=\_n\times\_H\@{R}+\dfrac{\_E\@{R}}{Z\@{R,\Pi}}.
\end{equation}
\end{subequations}
The equivalent boundary conditions for the whole $\Pi$-network can be established by substituting  the values of the inner tangential fields into the middle sheet boundary conditions of Eqs.~\eqref{eq:bound_cond_cpi}:
\begin{subequations}
\begin{equation}
\begin{split}
  \_n\times\left(\_E\@{R}-\_E\@{L}\right)=\dfrac{Z\@{C,\Pi}}{2}&\bigg[\_H\@{R}+\_H\@{L}.\\
  &\left.-\_n\times\left(\dfrac{\_E\@{R}}{Z\@{R,\Pi}}-\dfrac{\_E\@{L}}{Z\@{L,\Pi}}\right)\right], 
  \end{split}
\end{equation}
\begin{equation}
  \_n\times\left(\_H\@{R}-\_H\@{L}\right)=-\left(\dfrac{\_E\@{R}}{Z\@{R,\Pi}}+\dfrac{\_E\@{L}}{Z\@{L,\Pi}}\right).
\end{equation}
\end{subequations}



Using the definitions of the surface currents based on average fields of Eqs.~\eqref{eq:eq_curr_biani_model}, the coupling parameters for a $\Pi$-network read
\begin{subequations}
\label{eq:rho_pi_model}
\begin{equation}
  Y\@{ee,\Pi}=\dfrac{4\left(Z\@{L,\Pi}+Z\@{C,\Pi}+Z\@{R,\Pi}\right)}{Z\@{L,\Pi}Z\@{C,\Pi} + Z\@{C,\Pi}Z\@{R,\Pi} + 4Z\@{R,\Pi}Z\@{L,\Pi}}, 
\end{equation}
\begin{equation}
\begin{split}
  \chi\@{me,\Pi}&=\gamma\@{em,\Pi}\\
  &=\dfrac{2 Z\@{C,\Pi} \left(Z\@{R,\Pi}-Z\@{L,\Pi}\right)}{Z\@{L,\Pi}Z\@{C,\Pi} + Z\@{C,\Pi}Z\@{R,\Pi} + 4Z\@{R,\Pi}Z\@{L,\Pi}},
\end{split}
\end{equation}
\begin{equation}
  Z\@{mm,\Pi}=\dfrac{4 Z\@{R,\Pi} Z\@{C,\Pi}Z\@{L,\Pi}}{Z\@{L,\Pi}Z\@{C,\Pi} + Z\@{C,\Pi}Z\@{R,\Pi} + 4Z\@{R,\Pi}Z\@{L,\Pi}}.
\end{equation}
\end{subequations}

As shown before for the T-network, the relation $\chi\@{me,\Pi}=\gamma\@{em,\Pi}$ holds due to the reciprocal nature of the $\Pi$-network. Similarly as done in Eqs.~\eqref{eq:t_model_eq_impedances}, the impedances for a $\Pi$-network can be written in terms of the coupling parameters as
\begin{subequations}
\label{eq:pi_model_eq_impedances}
\begin{equation}
  Z\@{L,\Pi}=\dfrac{2 Z\@{mm,\Pi}}{Y\@{ee,\Pi}Z\@{mm,\Pi}+2\gamma\@{em,\Pi}+\gamma\@{em,\Pi}^2}, 
\end{equation}
\begin{equation}
  Z\@{C,\Pi}=\dfrac{4 Z\@{mm,\Pi}}{4-Y\@{ee,\Pi}Z\@{mm,\Pi}-\gamma\@{em,\Pi}^2},
\end{equation}
\begin{equation}
  Z\@{R,\Pi}=\dfrac{2 Z\@{mm,\Pi}}{Y\@{ee,\Pi}Z\@{mm,\Pi}-2\gamma\@{em,\Pi}+\gamma\@{em,\Pi}^2}.
\end{equation}
\end{subequations}
%

\renewcommand\thefigure{D.\arabic{figure}}    
\setcounter{figure}{0}   

\renewcommand\theequation{D\arabic{equation}}    
\setcounter{equation}{0}

\section*{APPENDIX D: Transformation between T and $\Pi$ models and their corresponding impedance and admittance matrices}

Due to the reciprocal nature of the system, these two models are fully interchangeable. Using the relations in Eqs.~\eqref{eq:rho_t_model} and \eqref{eq:rho_pi_model}, the corresponding impedances can be converted using relations
\begin{subequations}
\label{eq:t_to_pi}
\begin{equation}
  Z\@{L,\Pi}=\dfrac{Z\@{L,T}Z\@{C,T}+Z\@{C,T}Z\@{R,T}+Z\@{R,T}Z\@{L,T}}{Z\@{R,T}}, 
\end{equation}
\begin{equation}
  Z\@{C,\Pi}=\dfrac{Z\@{L,T}Z\@{C,T}+Z\@{C,T}Z\@{R,T}+Z\@{R,T}Z\@{L,T}}{Z\@{C,T}}, 
\end{equation}
\begin{equation}
  Z\@{R,\Pi}=\dfrac{Z\@{L,T}Z\@{C,T}+Z\@{C,T}Z\@{R,T}+Z\@{R,T}Z\@{L,T}}{Z\@{L,T}}; 
\end{equation}
\end{subequations}
and
\begin{subequations}
\label{eq:pi_to_t}
\begin{equation}
  Z\@{L,T}=\dfrac{Z\@{L,\Pi}Z\@{C,\Pi}}{Z\@{L,\Pi}+Z\@{C,\Pi}+Z\@{R,\Pi}}, 
\end{equation}
\begin{equation}
  Z\@{C,T}=\dfrac{Z\@{L,\Pi}Z\@{R,\Pi}}{Z\@{L,\Pi}+Z\@{C,\Pi}+Z\@{R,\Pi}}, 
\end{equation}
\begin{equation}
  Z\@{R,T}=\dfrac{Z\@{C,\Pi}Z\@{R,\Pi}}{Z\@{L,\Pi}+Z\@{C,\Pi}+Z\@{R,\Pi}}. 
\end{equation}
\end{subequations}

Alternatively, the T-network model parameters can be expressed in terms of a reciprocal impedance matrix (where $Z\@{12}=Z\@{21}$) \cite{Pozar_2004}. In that case, the impedances are given by
\begin{subequations}
\label{eq:t_to_z}
\begin{equation}
  Z\@{L,T}=Z\@{11}-Z\@{12}, 
\end{equation}
\begin{equation}
  Z\@{C,T}=Z\@{12},
\end{equation}
\begin{equation}
  Z\@{L,T}=Z\@{22}-Z\@{12};
\end{equation}
\end{subequations}
and the field coupling parameters read
\begin{subequations}
\label{eq:rho_t_z_model}
\begin{equation}
  Y\@{ee,Z}=\dfrac{4}{Z\@{11}+ 2Z\@{12} + Z\@{22}}, 
\end{equation}
\begin{equation}
  \chi\@{me,Z}=\gamma\@{em,Z}=\dfrac{2 \left( Z\@{22} - Z\@{11}\right)}{Z\@{11}+ 2Z\@{12} + Z\@{22}},
\end{equation}
\begin{equation}
  Z\@{mm,Z}=\dfrac{4 \left( Z\@{22} Z\@{11} - Z\@{12}^2\right)}{Z\@{11}+ 2Z\@{12} + Z\@{22}}.
\end{equation}
\end{subequations}
For the $\Pi$ model of Eqs.~\eqref{eq:rho_pi_model}, a reciprocal admittance matrix model (with $Y\@{12}=Y\@{21}$) is used instead \cite{Pozar_2004}. The impedances of the $\Pi$ model read in terms of the admittance matrix parameters as
\begin{subequations}
\label{eq:pi_to_y}
\begin{equation}
  Z\@{L,\Pi}=\dfrac{1}{Y\@{11}+Y\@{12}}, 
\end{equation}
\begin{equation}
  Z\@{C,\Pi}=-\dfrac{1}{Y\@{12}},
\end{equation}
\begin{equation}
  Z\@{L,\Pi}=\dfrac{1}{Y\@{22}+Y\@{12}};
\end{equation}
\end{subequations}
and the coupling parameters are expressed as
\begin{subequations}
\label{eq:rho_pi_y_model}
\begin{equation}
  Y\@{ee,Y}=\dfrac{4\left(Y\@{11}Y\@{22}-Y\@{12}^2\right)}{Y\@{11}- 2Y\@{12} + Y\@{22}}, 
\end{equation}
\begin{equation}
  \chi\@{me,Y}=\gamma\@{em,Y}=\dfrac{2 \left( Y\@{11} - Y\@{22}\right)}{Y\@{11}- 2Y\@{12} + Y\@{22}},
\end{equation}
\begin{equation}
  Z\@{mm,Y}=\dfrac{4 }{Y\@{11}- 2Y\@{12} + Y\@{22}}.
\end{equation}
\end{subequations}

With these two models, it becomes easier to design and characterize metasurfaces with asymmetric but reciprocal two-port responses.

\renewcommand\thefigure{E.\arabic{figure}}    
\setcounter{figure}{0}   

\renewcommand\theequation{E\arabic{equation}}    
\setcounter{equation}{0}

\section*{APPENDIX E: Reference-plane matching of a $\Delta\Sigma$ metasurface pair}
\begin{figure}[h]
    \centering    \includegraphics[width=1\linewidth]{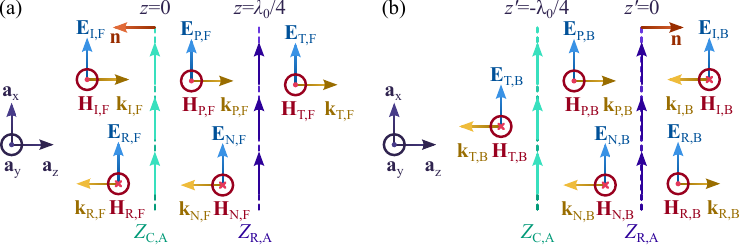}
    \caption{A $\Delta\Sigma$ metasurface pair can be analyzed using single illumination from (a) forward direction and (b) backward direction. In terms of phase, the reference plane is chosen to be at the first metasurface that is illuminated by the incident wave.}
    \label{fig:sche_ms_pair}
\end{figure}

Let us consider two cascaded metasurfaces, separated by $d=\lambda_0/4$ and illuminated by a single plane wave, as shown in Fig.~\ref{fig:sche_ms_pair}. The two metasurfaces can be modelled as two sheets supporting electric surface current  with surface impedances $Z\@{C,A}$ and $Z\@{R,A}$. In the case of Fig.~\ref{fig:sche_ms_pair}(a), the incident wave propagates in the $\_a\@{z}$ direction, and the coordinates origin is located at the first metasurface that the incident wave illuminates. The fields across the structure read
\begin{subequations}
\begin{equation}
    \begin{matrix}
    \_E\@{I,F}=E\@{F} e^{-j k_0 z} \_a\@{x}, &  \_H\@{I,F}=\dfrac{E\@{F}}{\eta_0} e^{-j k_0 z} \_a\@{y},
    \end{matrix}
\end{equation}
\begin{equation}
    \begin{matrix}
    \_E\@{T,F}=E\@{T,F} e^{-j k_0 z} \_a\@{x}, &  \_H\@{T,F}=\dfrac{E\@{T,F}}{\eta_0} e^{-j k_0 z} \_a\@{y},
    \end{matrix}
\end{equation}
\begin{equation}
    \begin{matrix}
    \_E\@{R,F}=E\@{R,F} e^{j k_0 z} \_a\@{x}, &  \_H\@{R,F}=-\dfrac{E\@{R,F}}{\eta_0} e^{j k_0 z} \_a\@{y},
    \end{matrix}
\end{equation}
\begin{equation}
    \begin{matrix}
    \_E\@{P,F}=E\@{P,F} e^{-j k_0 z} \_a\@{x}, &  \_H\@{P,F}=\dfrac{E\@{P,F}}{\eta_0} e^{-j k_0 z} \_a\@{y},
    \end{matrix}
\end{equation}
\begin{equation}
    \begin{matrix}
    \_E\@{N,F}=E\@{N,F} e^{j k_0 z} \_a\@{x}, &  \_H\@{N,F}=-\dfrac{E\@{N,F}}{\eta_0} e^{j k_0 z} \_a\@{y}.
    \end{matrix}
\end{equation}
\end{subequations}

The boundary conditions for the first metasurface at $z=0$ read
\begin{subequations}
\begin{equation}
    E\@{F}+E\@{R,F}=E\@{P,F}+E\@{N,F},
\end{equation}
\begin{equation}
\begin{split}
    \dfrac{1}{\eta_0}\left(E\@{F}-E\@{R,F}-E\@{P,F}+E\@{B,F} \right)&=J\@{C,A}\\
    &=\dfrac{E\@{F}+E\@{R,F}}{Z\@{C,A}};
\end{split}    
\end{equation}
\end{subequations}
and they are satisfied for the fields between metasurfaces
\begin{subequations}
\begin{equation}
    E\@{P,F}=-\dfrac{E\@{F}\left(\eta_0 - 2 Z\@{C,A}\right)+E\@{R,F} \eta_0}{2 Z\@{C,A}},
\end{equation}
\begin{equation}
    E\@{N,F}=\dfrac{E\@{F} \eta_0 + E\@{R,F}\left(\eta_0 + 2 Z\@{C,A}\right)}{2 Z\@{C,A}}.
\end{equation}
\label{eq:pair_forward_inner_fields}
\end{subequations}
Likewise, the boundary conditions for the second metasurface at $z=d$ read
\begin{subequations}
\begin{equation}
    E\@{P,F}e^{-j k_0 d}+E\@{N,F}e^{j k_0 d}=E\@{T,F}e^{-j k_0 d},
\end{equation}
\begin{equation}
\begin{split}
    \dfrac{1}{\eta_0}&\left(E\@{P,F}e^{-j k_0 d}-E\@{N,F}e^{j k_0 d}\right.\\
    &\left.-E\@{T,F}e^{-j k_0 d} \right)=J\@{R,A}=\dfrac{E\@{T,F}e^{-j k_0 d}}{Z\@{R,A}}.
    \end{split}
\end{equation}
\label{eq:pair_forward_second_boundary}
\end{subequations}
By combining  Eqs.~\eqref{eq:pair_forward_inner_fields} with  Eqs.~\eqref{eq:pair_forward_second_boundary}, it can be found that the amplitudes of scattered waves produced by the cascaded metasurfaces (assuming $Z\@{R,A}=Z\@{C,A}$) read as in Eqs.~\eqref{eq:pair_forward_outer_fields}.
%

The complementary scenario of Fig. \ref{fig:sche_ms_pair}(b) defines  the fields produced under backward illumination, where the first illuminated metasurface is  the coordinate reference ($z'=0$):
\begin{subequations}
\begin{equation}
    \begin{matrix}
    \_E\@{I,B}=E'\@{B} e^{j k_0 z'} \_a\@{x}, &  \_H\@{I,B}=-\dfrac{E'\@{B}}{\eta_0} e^{j k_0 z'} \_a\@{y},
    \end{matrix}
\end{equation}
\begin{equation}
    \begin{matrix}
    \_E\@{T,B}=E'\@{T,B} e^{j k_0 z'} \_a\@{x}, &  \_H\@{T,B}=-\dfrac{E'\@{T,B}}{\eta_0} e^{j k_0 z'} \_a\@{y},
    \end{matrix}
\end{equation}
\begin{equation}
    \begin{matrix}
    \_E\@{R,B}=E'\@{R,B} e^{-j k_0 z'} \_a\@{x}, &  \_H\@{R,B}=\dfrac{E'\@{R,B}}{\eta_0} e^{-j k_0 z'} \_a\@{y},
    \end{matrix}
\end{equation}
\begin{equation}
    \begin{matrix}
    \_E\@{P,B}=E'\@{P,B} e^{-j k_0 z'} \_a\@{x}, &  \_H\@{P,B}=\dfrac{E'\@{P,B}}{\eta_0} e^{-j k_0 z'} \_a\@{y},
    \end{matrix}
\end{equation}
\begin{equation}
    \begin{matrix}
    \_E\@{N,B}=E'\@{N,B} e^{j k_0 z'} \_a\@{x}, &  \_H\@{N,B}=-\dfrac{E'\@{N,B}}{\eta_0} e^{j k_0 z'} \_a\@{y}.
    \end{matrix}
\end{equation}
\end{subequations}
Therefore, the boundary conditions at the metasurface located at $z'=0$ read
\begin{subequations}
\begin{equation}
    E'\@{B}+E'\@{R,B}=E'\@{P,B}+E'\@{N,B},
\end{equation}
\begin{equation}
\begin{split}
    \dfrac{1}{\eta_0}\left(E'\@{I,B}-E'\@{R,B}-E'\@{N,B}+E\@{P,B} \right)&=J\@{R,A}\\
    &=\dfrac{E'\@{I,B}+E'\@{R,B}}{Z\@{R,A}}.
\end{split}    
\end{equation}
\end{subequations}
Similarly as found for the forward illumination, the inner fields read
\begin{subequations}
\begin{equation}
    E'\@{P,B}=\dfrac{E'\@{B} \eta_0+E'\@{R,B}\left(\eta_0 + 2 Z\@{R,A}\right)}{2 Z\@{R,A}},
\end{equation}
\begin{equation}
    E'\@{N,B}=-\dfrac{E'\@{B}\left(\eta_0 - 2 Z\@{R,A}\right) + E'\@{R,B} \eta_0 }{2 Z\@{R,A}}.
\end{equation}
\label{eq:pair_backward_inner_fields}
\end{subequations}
The boundary conditions at the second metasurface at $z'=-d$ read
\begin{subequations}
\begin{equation}
    E'\@{P,B}e^{-j k_0 d}+E'\@{N,B}e^{j k_0 d}=E'\@{T,B}e^{-j k_0 d},
\end{equation}
\begin{equation}
\begin{split}
    \dfrac{1}{\eta_0}&\left(E'\@{N,B}e^{-j k_0 d}-E'\@{P,B}e^{j k_0 d}\right.\\
    &\left.-E'\@{T,B}e^{-j k_0 d} \right)=J\@{C,A}=\dfrac{E'\@{T,B}e^{-j k_0 d}}{Z\@{C,A}}.
\end{split}    
\end{equation}
\label{eq:pair_forward_second_boundary_prime}
\end{subequations}
The combination of both boundary conditions with the same surface impedance $Z\@{R,A}=Z\@{C,A}$ is  satisfied for scattered waves
\begin{subequations}
\begin{equation}
    E'\@{T,B}=\dfrac{2 E'\@{B} Z\@{C,A}^2}{\eta_0^2+2 \eta_0 Z\@{C,A} + 2 Z\@{C,A}^2},
\end{equation}
\begin{equation}
    E'\@{R,B}=-\dfrac{ E'\@{B} \eta_0^2}{\eta_0^2+2 \eta_0 Z\@{C,A} + 2 Z\@{C,A}^2}.
\end{equation}
\label{eq:pair_backward_outer_fields}
\end{subequations}

In order to combine the scattering produced by the two incident waves  ($\_E\@{R,F}$ with $\_E\@{T,B}$ and $\_E\@{R,B}$ with $\_E\@{T,F}$, respectively), both scattered fields should refer to the same reference plane (we place it at the metasurface that is illuminated first by the forward incident wave). This is done using the relation $z'=z-d$, with the fields redefined as
\begin{subequations}
\begin{equation}
    \_E\@{I,B}=E\@{B} e^{j k_0 z} \_a\@{x} \rightarrow E\@{I,B}=E'\@{B} e^{-j k_0 d},
\end{equation}
\begin{equation}
    \_E\@{T,B}=E\@{T,B} e^{j k_0 z} \_a\@{x} \rightarrow E\@{T,B}=E'\@{T,B} e^{-j k_0 d},
\end{equation}
\begin{equation}
    \_E\@{R,B}=E\@{R,B} e^{-j k_0 z} \_a\@{x} \rightarrow E\@{R,B}=E'\@{R,B} e^{j k_0 d}.
\end{equation}
\label{eq:pair_backward_shift}
\end{subequations}
This quarter-wavelength shift of the reference plane affects drastically the phase relation between the backward wave and its corresponding reflected wave, without affecting the phase in transmission. This property can be noticed by replacing the results of Eqs.~\eqref{eq:pair_backward_shift} into the transfer functions of Eqs.~\eqref{eq:pair_backward_outer_fields}, which can be rewritten as
\begin{subequations}
\begin{equation}
    E\@{T,B}=E'\@{T,B}=\dfrac{2 E\@{B} Z\@{C,A}^2}{\eta_0^2+2 \eta_0 Z\@{C,A} + 2 Z\@{C,A}^2},
\end{equation}
\begin{equation}
    E\@{R,B}=e^{2j k_0 d}E'\@{R,B}=\dfrac{ E\@{B} \eta_0^2}{\eta_0^2+2 \eta_0 Z\@{C,A} + 2 Z\@{C,A}^2}.
\end{equation}
\label{eq:pair_backward_outer_fields_prime}
\end{subequations}
This combination of asymmetrically-defined reference plane in a structure with a  significant electrical distance between metasurfaces introduces the phase-shift in reflection for backward illumination required for realization of a  $\Delta\Sigma$ metasurface pair.

\renewcommand\thefigure{F.\arabic{figure}}    
\setcounter{figure}{0}   

\renewcommand\theequation{F\arabic{equation}}    
\setcounter{equation}{0}

\section*{APPENDIX F: Performance degradation due to angular mismatch}

In the scenario that one of the incident sources is not properly aligned to the $\Delta\Sigma$ metasurface pair
\begin{subequations}
\label{eq:mismatched_phase_fow}
\begin{equation}
 \_E\@{I,F}=E\@{F} e^{j \Delta\Phi} e^{-j k_0 z} \_a\@{x}, 
\end{equation}
\begin{equation}
 \_E\@{I,B}=E\@{B} e^{j k_0 z} \_a\@{x}, 
\end{equation}
\end{subequations}
the shifted phase in the forward component $\Delta\Phi$ will be replicated into the scattering waves
\begin{subequations}
\label{eq:mismatched_phase_inc}
\begin{equation}
 \_E\@{\Sigma}=\alpha \left(E\@{F}e^{j \Delta\Phi} + E\@{B} \right)e^{j k_0 z} \_a\@{x}, 
\end{equation}
\begin{equation}
 \_E\@{\Delta}=\alpha \left(E\@{F}e^{j \Delta\Phi} - E\@{B} \right) e^{-j k_0 z} \_a\@{x}.
\end{equation}
\end{subequations}
In the scenario that both incident waves have the same amplitude and phase $E\@{F}=E\@{B}$, the sum and difference components can be rewritten as
\begin{subequations}
\label{eq:mismatched_phase_scatt}
\begin{equation}
 \_E\@{\Sigma}=\alpha E\@{F}\left(e^{j \Delta\Phi} + 1 \right)e^{j k_0 z} \_a\@{x}, 
\end{equation}
\begin{equation}
 \_E\@{\Delta}=\alpha E\@{F} \left(e^{j \Delta\Phi} - 1 \right) e^{-j k_0 z} \_a\@{x}.
\end{equation}
\end{subequations}
Ideally, the incident wave propagating in the forward direction does not have any mismatch $\Delta\Phi=0$, and the sum component has an amplitude $E\@{\Sigma}=2\alpha E\@{F}$, while the difference component fades $E\@{\Delta}=0$. With this information, it is possible to estimate the attenuation produced by the phase mismatch in terms of the scattered power in the sum component
\begin{equation}
    L_{\Delta\Phi}=\dfrac{\left\vert \alpha E\@{F}\left(e^{j \Delta\Phi} + 1 \right) \right\vert^2}{\left\vert 2\alpha E\@{F} \right\vert^2}=\dfrac{1+\cos \Delta\Phi}{2}.\label{eq:phase_attenuation}
\end{equation}
From Eq.~\eqref{eq:phase_attenuation} it can be found that the scattered power decays down to 95\% of the ideal value ($L_{\Delta\Phi}=0.95$) when the phase mismatch becomes $\Delta\Phi=\pm 25.84^\circ$.

\renewcommand\thefigure{G.\arabic{figure}}    
\setcounter{figure}{0}   

\renewcommand\theequation{G\arabic{equation}}    
\setcounter{equation}{0}

\section*{APPENDIX G: Scattering produced by an bianisotropic sheet with uniaxial response under elliptical illumination}

Let us consider the same bianisotropic sheet of Fig.\ref{fig:sche_biani} and illuminated it with an electromagnetic wave on the form
\begin{subequations}
\begin{equation}
    \_E\@{I,F}=\left(E\@{F,x}\_a\@{x} + E\@{F,y}\_a\@{y} \right)e^{-j k_0 z} ,
\end{equation}
\begin{equation}
    \_H\@{I,F}=\dfrac{1}{\eta_0}\left(- E\@{F,y}\_a\@{x} + E\@{F,x}\_a\@{y} \right)e^{-j k_0 z},
\end{equation}
\end{subequations}
where $E\@{F,x}$ ($E\@{F,y}$) is the complex amplitude of the electric field projected onto the $x$ ($y$)-axis. The relation between the $x$ and $y$ components of the electric field defines the wave properties (linear, circular or elliptical) and its handedness (left-handed or right-handed). On the presence of such particular incident wave, the produced scattering waves acquires a similar shape

\begin{subequations}
\begin{equation}
    \_E\@{T,F}=\left(E\@{TF,x}\_a\@{x} + E\@{TF,y}\_a\@{y} \right)e^{-j k_0 z} ,
\end{equation}
\begin{equation}
    \_H\@{T,F}=\dfrac{1}{\eta_0}\left( - E\@{TF,y}\_a\@{x} +E\@{TF,x}\_a\@{y} \right)e^{-j k_0 z},
\end{equation}
\begin{equation}
    \_E\@{R,F}=\left(E\@{RF,x}\_a\@{x} + E\@{RF,y}\_a\@{y} \right)e^{j k_0 z} , 
\end{equation}
\begin{equation}
    \_H\@{R,F}=\dfrac{1}{\eta_0}\left( E\@{RF,y}\_a\@{x} - E\@{TR,x}\_a\@{y} \right)e^{j k_0 z}.
\end{equation}
\end{subequations}

By solving the boundary conditions of Eqs.~\eqref{eq:eq_curr_biani_model}, it can be found that for a bianisotropic sheet with uniaxial response the amplitude of the scattering waves only depends on the corresponding coplanar component of the incident field, resulting in similar expressions found in Eqs.~\eqref{eq:eq_bound_biani_forward}
\begin{subequations}
\label{eq:eq_bound_elliptical_forward}
\begin{equation}
\begin{split}
  &-\left(\begin{bmatrix} E\@{TF,x} \\ E\@{TF,y} \end{bmatrix} -\begin{bmatrix} E\@{F,x} \\ E\@{F,y} \end{bmatrix}-\begin{bmatrix} E\@{RF,x} \\ E\@{RF,y} \end{bmatrix}\right)\\
  &=\dfrac{Z\@{mm}}{2\eta_0}\left(\begin{bmatrix} E\@{TF,x} \\ E\@{TF,y} \end{bmatrix}+\begin{bmatrix} E\@{F,x} \\ E\@{F,y} \end{bmatrix}-\begin{bmatrix} E\@{RF,x} \\ E\@{RF,y} \end{bmatrix}\right)\\
  &-\dfrac{\gamma\@{em}}{2}\left(\begin{bmatrix} E\@{TF,x} \\ E\@{TF,y} \end{bmatrix}+\begin{bmatrix} E\@{F,x} \\ E\@{F,y} \end{bmatrix}+\begin{bmatrix} E\@{RF,x} \\ E\@{RF,y} \end{bmatrix}\right),
\end{split}  
\end{equation}
\begin{equation}
\begin{split}
  &\dfrac{1}{\eta_0}\left(\begin{bmatrix} E\@{TF,x} \\ E\@{TF,y} \end{bmatrix}-\begin{bmatrix} E\@{F,x} \\ E\@{F,y} \end{bmatrix}+\begin{bmatrix} E\@{RF,x} \\ E\@{RF,y} \end{bmatrix}\right)\\
  &=-\dfrac{Y\@{ee}}{2}\left(\begin{bmatrix} E\@{TF,x} \\ E\@{TF,y} \end{bmatrix}+\begin{bmatrix} E\@{F,x} \\ E\@{F,y} \end{bmatrix}+\begin{bmatrix} E\@{RF,x} \\ E\@{RF,y} \end{bmatrix}\right)\\
  &-\dfrac{\chi\@{me}}{2\eta_0}\left(\begin{bmatrix} E\@{TF,x} \\ E\@{TF,y} \end{bmatrix}+\begin{bmatrix} E\@{F,x} \\ E\@{F,y} \end{bmatrix}-\begin{bmatrix} E\@{RF,x} \\ E\@{RF,y} \end{bmatrix}\right).
\end{split}  
\end{equation}
\end{subequations}
As a result, the amplitudes of the transmitted wave read
\begin{equation}
\begin{bmatrix}
    E\@{TF,x} \\
    E\@{TF,y}
\end{bmatrix} =  \tau\@{F}
\begin{bmatrix}
    E\@{F,x} \\
    E\@{F,y}
\end{bmatrix},
\end{equation}
while the reflected ones are expressed as
\begin{equation}
\begin{bmatrix}
    E\@{RF,x} \\
    E\@{RF,y}
\end{bmatrix} =  \Gamma\@{F}
\begin{bmatrix}
    E\@{F,x} \\
    E\@{F,y}
\end{bmatrix};
\end{equation}
with the definitions of $\tau\@{F}$ and $\Gamma\@{F}$ found in Eqs.~\eqref{eq:coupling_parameters_forward}. 

This analysis can be also performed for a incident backward wave
\begin{subequations}
\begin{equation}
    \_E\@{I,B}=\left(E\@{B,x}\_a\@{x} + E\@{B,y}\_a\@{y} \right)e^{j k_0 z} ,
\end{equation}
\begin{equation}
    \_H\@{I,B}=\dfrac{1}{\eta_0}\left( E\@{B,y}\_a\@{x} - E\@{B,x}\_a\@{y} \right)e^{j k_0 z},
\end{equation}
\end{subequations}
which produces scattering waves
\begin{subequations}
\begin{equation}
    \_E\@{T,B}=\left(E\@{TB,x}\_a\@{x} + E\@{TB,y}\_a\@{y} \right)e^{j k_0 z},
\end{equation}
\begin{equation}
    \_H\@{T,B}=\dfrac{1}{\eta_0}\left( E\@{TB,y}\_a\@{x} -E\@{TB,x}\_a\@{y} \right)e^{j k_0 z},
\end{equation}
\begin{equation}
    \_E\@{R,B}=\left(E\@{RB,x}\_a\@{x} + E\@{RB,y}\_a\@{y} \right)e^{-j k_0 z}, 
\end{equation}
\begin{equation}
    \_H\@{R,B}=\dfrac{1}{\eta_0}\left( - E\@{RB,y}\_a\@{x} +E\@{RB,x}\_a\@{y} \right)e^{-j k_0 z}.
\end{equation}
\end{subequations}
The relation between incident and scattering waves is solved again through Eqs.~\eqref{eq:eq_curr_biani_model}, taking a form comparable to Eqs.~\eqref{eq:eq_bound_biani_backward}:
\begin{subequations}
\label{eq:eq_bound_elliptical_backward}
\begin{equation}
\begin{split}
  &\left(\begin{bmatrix} E\@{B,x} \\ E\@{B,y} \end{bmatrix}+\begin{bmatrix} E\@{RB,x} \\ E\@{RB,y} \end{bmatrix}-\begin{bmatrix} E\@{TB,x} \\ E\@{TB,y} \end{bmatrix}\right)\\
  &=\dfrac{Z\@{mm}}{2\eta_0}\left(\begin{bmatrix} E\@{B,x} \\ E\@{B,y} \end{bmatrix}-\begin{bmatrix} E\@{RB,x} \\ E\@{RB,y} \end{bmatrix}+\begin{bmatrix} E\@{TB,x} \\ E\@{TB,y} \end{bmatrix}\right)\\
  &+\dfrac{\gamma\@{em}}{2}\left(\begin{bmatrix} E\@{B,x} \\ E\@{B,y} \end{bmatrix}+\begin{bmatrix} E\@{RB,x} \\ E\@{RB,y} \end{bmatrix}+\begin{bmatrix} E\@{TB,x} \\ E\@{TB,y} \end{bmatrix}\right),
  \end{split}
\end{equation}
\begin{equation}
\begin{split}
  &\dfrac{1}{\eta_0}\left(\begin{bmatrix} E\@{B,x} \\ E\@{B,y} \end{bmatrix}-\begin{bmatrix} E\@{RB,x} \\ E\@{RB,y} \end{bmatrix}-\begin{bmatrix} E\@{TB,x} \\ E\@{TB,y} \end{bmatrix}\right)\\
  &=\dfrac{Y\@{ee}}{2}\left(\begin{bmatrix} E\@{B,x} \\ E\@{B,y} \end{bmatrix}+\begin{bmatrix} E\@{RB,x} \\ E\@{RB,y} \end{bmatrix}+\begin{bmatrix} E\@{TB,x} \\ E\@{TB,y} \end{bmatrix}\right)\\
  &-\dfrac{\chi\@{me}}{2\eta_0}\left(\begin{bmatrix} E\@{B,x} \\ E\@{B,y} \end{bmatrix}-\begin{bmatrix} E\@{RB,x} \\ E\@{RB,y} \end{bmatrix}+\begin{bmatrix} E\@{TB,x} \\ E\@{TB,y} \end{bmatrix}\right).
  \end{split}
\end{equation}
\end{subequations}
Hence, transmission and reflection coefficients for backward illumination follow the relation
\begin{equation}
\begin{bmatrix}
    E\@{TB,x} \\
    E\@{TB,y}
\end{bmatrix} =  \tau\@{B}
\begin{bmatrix}
    E\@{B,x} \\
    E\@{B,y}
\end{bmatrix},
\end{equation}
\begin{equation}
\begin{bmatrix}
    E\@{RB,x} \\
    E\@{RB,y}
\end{bmatrix} =  \Gamma\@{B}
\begin{bmatrix}
    E\@{B,x} \\
    E\@{B,y}
\end{bmatrix};
\end{equation}
with $\tau\@{B}$ and $\Gamma\@{B}$ taken from Eqs.~\eqref{eq:coupling_parameters_backward}. Please notice that the reflected wave for both cases (forward and backward illumination) has the opposite handedness compared to the incident source. While the electromagnetic fields at the interface are proportional to the ones of the incident wave, their wavevectors are pointing at opposite directions.

%


\end{document}